\documentclass[rotating,dvips]{arxstspdf}
\usepackage{listings}
\usepackage{dcolumn}
\usepackage{graphicx}
\usepackage{flushend}
\usepackage{stfloats}

\volume{25}
\issue{3}
\pubyear{2010}
\firstpage{325}
\lastpage{347}
\doi{10.1214/10-STS328}

\makeatletter

\newcolumntype{d}[1]{D{.}{.}{#1}}
\DeclareMathAlphabet\mathcaligr{OMS}{cmsy}{m}{n}
\renewcommand{\mathcal}{\mathcaligr}
\newcommand{\cal}{\mathcaligr}
\newcommand{\bbi}{\mathbf{b}_i}
\newcommand{\Yi}{\mathbf{Y}_i}
\renewcommand{\bm}{\mathbf}
\newcommand{\bi}{\mathbf{b}_i}

\newcommand{\rE}{E}
\newcommand{\bfzeta}{\bolds\zeta}
\newcommand{\BY}{\mathbf{Y}}
\newcommand{\bx}{\mathbf{x}}
\newcommand{\bz}{\mathbf{z}}

\newcommand{\bfSigma}{\bolds\Sigma}
\newcommand{\bftheta}{\bolds\theta}
\newcommand{\bfvartheta}{\bolds\vartheta}
\newcommand{\bfmu}{\bolds\mu}

\newcommand{\Var}{\operatorname{Var}}
\newcommand{\Y}{\mathbf{Y}}
\newcommand{\bfbeta}{\bolds\xi}
\newcommand{\Corr}{\operatorname{Corr}}
\makeatother

\begin{document}
\begin{frontmatter}

\title{A Family of Generalized Linear Models for Repeated Measures with
Normal and Conjugate Random Effects}
\runtitle{Models With Normal and Conjugate Random Effects}

\begin{aug}
\author[a]{\fnms{Geert} \snm{Molenberghs}\corref{}\ead[label=e1]{geert.molenberghs@uhasselt.be}},
\author[b]{\fnms{Geert} \snm{Verbeke}\ead[label=e2]{geert.verbeke@med.kuleuven.be}},
\author[c]{\fnms{Clarice G. B.} \snm{Dem\'etrio}\ead[label=e3]{clarice@esalq.usp.br}}
\and
\author[d]{\fnms{Afr\^anio M. C.} \snm{Vieira}\ead[label=e4]{fran.usp@gmail.com}}
\runauthor{Molenberghs, Verbeke, Dem\'etrio and Vieira}

\affiliation{Universiteit Hasselt, Katholieke Universiteit Leuven,
ESALQ and Universidade de Bras\'\i lia}
\vspace{-3pt}
\address[a]{Geert Molenberghs is Professor of Biostatistics, I-BioStat, Universiteit
Hasselt, B-3590 Diepenbeek, Belgium and
I-BioStat, Katholieke Universiteit Leuven, B-3000 Leuven, Belgium (\printead{e1}).}
\address[b]{Geert Verbeke is Professor of Biostatistics, I-BioStat, Katholieke
Universiteit Leuven, B-3000 Leuven, Belgium and
I-BioStat, Universiteit Hasselt, B-3590 Diepenbeek, Belgium (\printead{e2}).}
\address[c]{Clarice G. B. Dem\'etrio is Professor of Biometry, ESALQ,
Piracicaba, Sa\~o Paulo, Brazil (\printead{e3}).}
\address[d]{Afr\^anio M. C. Vieira is Professor of Statistics, ESALQ, Piracicaba,
Sa\~o Paulo, Brazil and
Universidade de Bras\'\i lia, Department de Estat\'\i stica, Bras\'\i
lia/DF, Brasil (\printead{e4}).}

\end{aug}

% ABSTRACT
%
\begin{abstract}
Non-Gaussian outcomes are often modeled using members of the so-called
exponential family. Notorious members are the Bernoulli model for
binary data, leading to logistic regression, and the Poisson model for
count data, leading to Poisson regression. Two of the main reasons for
extending this family are (1) the occurrence of overdispersion, meaning
that the variability in the data is not adequately described by the
models, which often exhibit a prescribed mean--variance link, and (2)
the accommodation of hierarchical structure in the data, stemming from
clustering in the data which, in turn, may result from repeatedly
measuring the outcome, for various members of the same family, etc. The
first issue is dealt with through a variety of overdispersion models,
such as, for example, the beta-binomial model for grouped binary data
and the negative-binomial model for counts. Clustering is often
accommodated through the inclusion of random subject-specific effects.
Though not always, one conventionally assumes such random effects to be
normally distributed. While both of these phenomena may occur
simultaneously, models combining them are uncommon. This paper proposes
a broad class of generalized linear models accommodating overdispersion
and clustering through two separate sets of random effects. We place
particular emphasis on so-called conjugate random effects at the level
of the mean for the first aspect and normal random effects embedded
within the linear predictor for the second aspect, even though our
family is more general. The binary, count and time-to-event cases are
given particular emphasis. Apart from model formulation, we present an
overview of estimation methods, and then settle for maximum likelihood
estimation with analytic--numerical integration. Implications for the
derivation of marginal correlations functions are discussed. The
methodology is applied to data from a study in epileptic seizures, a
clinical trial in toenail infection named onychomycosis and survival
data in children with asthma.
\end{abstract}

% KEYWORDS
%

\end{frontmatter}

%s1 ###
\section{Introduction}\label{intro}

Next to continuous outcomes, binary and binomial outcomes, counts and
times to event take a prominent place in applied modeling and the
corresponding methodological literature. It is common to place such
models within the generalized linear modeling (GLM) framework (Nelder
and Wedderburn, \citeyear{Nelder1972}; McCullagh and Nelder, \citeyear{McCullagh1989}; Agresti, \citeyear{Agresti2002}). This
framework allows one to restrict specification to first and second
moments only, on the one hand, or to fully formulate distributional
assumptions, on the other hand. When the latter route is chosen, the
exponential family (McCullagh and Nelder, \citeyear{McCullagh1989}) provides an elegant and
encompassing mathematical framework, because it has the normal,
Bernoulli/binomial, Poisson and Weibull/exponential models as prominent members.

The elegance of the framework draws from certain linearity properties
of the log-likelihood function, producing mathematically convenient
score equations and ultimately convenient-in-use inferential
instruments, both in terms of point and interval estimation as well as
for hypothesis testing.

Nevertheless, it has been clear for several decades, for binomial,
count and time-to-event data, that a key feature of the GLM framework
and many of the exponential family members, the so-called \textit{mean--variance relationship}, may be overly restrictive. By this
relationship, we indicate that the variance is a deterministic function
of the mean. For example, for Bernoulli outcomes with success
probability $\mu=\pi$, the variance is $v(\mu)=\pi(1-\pi)$, for
counts $v(\mu)=\mu$ and for the exponential model $v(\mu)=\mu^2$.
In contrast, for continuous, normally distributed outcomes, the mean
and variance are entirely separate parameters. While i.i.d. binary
data cannot contradict the mean--variance relationship, i.i.d. binomial
data,\break counts and survival data can. This explains why early work has
been devoted to formulating models that explicitly allow for
overdispersion or, more generally, to proposing models that enjoy less
restrictive mean--variance relationships. For purely binary data,
hierarchies need to be present in the data in order to violate the
mean--variance link. One such class of hierarchies is with repeated
measures or longitudinal data, where an outcome on a study subject is
recorded repeatedly over time. With such models gaining momentum, not
only for the Gaussian case (Verbeke and Molenberghs, \citeyear{Verbeke2000}), but also
for non-Gaussian data (Molenberghs and Verbeke, \citeyear{Molenberghs2005}),\vadjust{\goodbreak} extensions of
the GLM framework have been formulated. For other types of outcomes,
such hierarchical settings further compound the issue of overly
restrictive variance relationships. In all cases, hierarchies induce
association. These features taken together call for very flexible
models, doing proper justice to each of the mean, variance and
association structures.

Hinde and Dem\'etrio (\citeyear{Hinde1998a,Hinde1998b}) provide broad overviews of approaches for
dealing with overdispersion, considering moment-based as well as
full-distribution avenues. Placing most emphasis on the \mbox{binomial} and
Poisson settings, they pay particular attention to random-effects-based
solutions to the problem, including but not limited to the
beta-binomial model (Skellam, \citeyear{Skellam1948}; Kleinman, \citeyear{Kleinman1973}) for binary and
binomial data and with beta random effects, and the negative-binomial
model (Breslow, \citeyear{Breslow1984}; Lawless, \citeyear{Lawless1987}), where the natural parameter is
assumed to follow a gamma distribution. The said gamma distribution
also features in many so-called frailty models, that is, specific
random-effects models for time-to-event data (Duchateau and Janssen,
\citeyear{Duchateau2007}). On the other hand, especially focusing on hierarchical data, the
so-called generalized linear mixed model (GLMM, Engel and Keen, \citeyear{Engel1994};
Breslow and Clayton, \citeyear{Breslow1993}; Wolfinger and O'Connell, \citeyear{Wolfinger1993}) has gained
popularity as a tool to accommodate overdispersion and/or
hierarchy-induced association for outcomes that are not necessarily of
a Gaussian type, in spite of problems, not only of a computational
type, but also in terms of interpretation. These arise from the
combination of general exponential family models with normally
distributed random effects. Unlike for Gaussian data, the derivation of
marginal moments and joint distributions is less than straightforward,
even though in this paper we make progress beyond what is available in
the literature. Part of GLMMs popularity originates from the
availability of implementations in a variety of standard software
packages. Other solutions to accommodating overdispersion include
mixture modeling and specific models for zero-inflated Poisson models
(Ridout, Dem\'etrio and Hinde, \citeyear{Ridout1998}; B\"ohning, \citeyear{B2000};\break
McLachlan and
Peel, \citeyear{McLachlan2000}).

Important unifying and computational progress has been made by Lee and
Nelder (\citeyear{Lee1996,Lee2001a,Lee2001b,Lee2003}) (see also Lee, Nelder and Pawitan, \citeyear{Lee2006}) by
proposing so-called \textit{hierarchical generalized linear models},
offering a broad class of outcome and random-effects distributions,
combined with appealing computational schemes. Unification has also
been reached by Skrondal and Rabe-Hesketh (\citeyear{Skrondal2004}), who assemble under
the same roof a number of modeling\vadjust{\goodbreak} strands,  such as multilevel
modeling, structural equations modeling, latent variables, latent
classes and random-effects models for longitudinal and otherwise
hierarchical data.

In this paper we introduce a general and flexible framework for such
combinations, starting from arbitrary generalized linear models and
exponential family members. Specific emphasis is placed on normally
distributed, binary, binomial, count and time-to-event outcomes. There
are various reasons to do so. First, non-Gaussian hierarchical data
exhibit three important features: (1) the mean structure; (2) the
variance structure; and (3) the correlation structure. Our proposed
framework features: (a) a mean structure; (b) overdispersion, often
conjugate random-effects; (c) normal random effects. It will be clear
from our case studies that model fit can be improved, and hence model
interpretation changed, by shifting to the extended model. Second,
especially in cases where the variance and/or correlation structures
are of interest (e.g., surrogate marker evaluation, psychometric
evaluation, etc.), such extensions are useful. Third, even when
interest remains with more conventional models, such as the GLMM, the
extended model can serve as a goodness-of-fit tool. Fourth, because we
can derive closed-form expressions for both standard and extended
models, the accuracy of parameter estimation and resulting inferences
can be improved, while obviating the need for tedious numerical
integration techniques. Fifth, the analysis of the case studies
corroborates this need. Such needs were recognized by Booth et al. (\citeyear{Booth2003}) and
Molenberghs, Verbeke and Dem\'etrio (\citeyear{MolenberghsDem2007}) who, in
the context of count data, formulated a model combining normal and
gamma random effects.

The paper is organized as follows. In Section~\ref{case studies} three
motivating case studies, with binary data, counts and survival data,
respectively, are described, with analyses reported near the end of the
manuscript, in Section~\ref{analysiscasestudies}. Basic ingredients
for our modeling framework, standard generalized linear models,
extensions for overdispersion and the generalized linear mixed model,
are the subject of Section~\ref{ingredients}. The proposed, combined
model is described and further studied in Section~\ref{combined}.
Avenues for parameter estimation and ensuing inferences are explored in
Section~\ref{estimation}. There are several appendices. Supplementary Material~A
offers generic approximations for means and variances. Supplementary Material~B--E
provide details for the Poisson case, the binary case with logit link
and the binary case with probit link, and the time-to-event case,
respectively. Implications of our findings for the derivation of
marginal correlation functions are the topic of Section~F in the Supplementary Material.

%s2 ###
\section{Case Studies}\label{case studies}

We will describe three case studies. The first one producing count
data, the second one with binary data, and the third one of a
time-to-event type.

%s2.1 ###
\subsection{A Clinical Trial in Epileptic Patients}\label{dataepilepsy}

The data considered here are obtained from a randomized,
double-blind, parallel group multi-center study for the comparison of
placebo with a new anti-epileptic drug (AED), in combination with one
or two other AED's. The study is described in full detail in Faught
et al. (\citeyear{Faught1996}). The randomization of epilepsy patients took
place after a 12-week baseline period that served as a stabilization
period for the use of AED's, and during which the number of seizures
were counted. After that period, $45$ patients were assigned to the
placebo group and $44$ to the active (new) treatment group. Patients
were then measured weekly. Patients were followed (double-blind) during
16 weeks, after which they were entered into a long-term open-extension
study. Some patients were followed for up to 27 weeks. The outcome of
interest is the number of epileptic seizures experienced during the
most recent week. The research question is whether or not the
additional new treatment reduces the number of epileptic seizures.

%s2.2 ###
\subsection{A Casfe Study in Onychomycosis}\label{dataonychomycosis}

These data come from a randomized, double-blind, parallel group,
multicenter study for the comparison of two oral
treatments (coded as $A$ and $B$) for toenail dermatophyte
onychomycosis (TDO), described in full detail by De Backer et al. (\citeyear{De1996}).
TDO is a common toenail infection, difficult to treat,
affecting more than 2 out of 100 persons (Roberts et al., \citeyear{Roberts1992}).
Anti-fungal compounds, classically used for treatment of TDO, need to
be taken
until the whole nail has grown out healthy. The development of new such
compounds, however, has reduced the treatment duration to 3 months. The
aim of the present study was to compare the efficacy and safety of 12
weeks of continuous therapy with treatment $A$ or with treatment $B$.
In total, $2\times189$ patients, distributed over 36 centers, were
randomized. Subjects were followed during 12 weeks (3~months) of
treatment and followed further, up to a total of 48 weeks (12 months).
Measurements were taken at baseline, every month during treatment and
every 3 months afterward, resulting in a maximum of 7 measurements per
subject. At the first occasion, the treating physician indicates one of
the affected toenails as the target nail, the nail which will be
followed over time. We will restrict our analyses to only those
patients for which the target nail was one of the two big toenails (146
and 148 subjects, in group A and group B, respectively). One of the
responses of interest was the unaffected nail length, measured from the
nail bed to the infected part of the nail, which is always at the free
end of the nail, expressed in mm. This outcome has been studied
extensively in Verbeke and Molenberghs (\citeyear{Verbeke2000}). Another important
outcome in this study was the severity of the infection, coded as 0
(not severe) or 1 (severe). The question of interest was whether the
percentage of severe infections decreased over time, and whether that
evolution was different for the two treatment groups.

%s2.3 ###
\subsection{Recurrent Asthma Attacks in Children} \label{datasurvival}

These data have been studied in Duchateau and Janssen (\citeyear{Duchateau2007}). Asthma is
occurring more and more frequently in very young children (between 6
and 24~months). Therefore, a new application of an existing
anti-allergic drug is administered to children who are at higher risk
to develop asthma in order to prevent it. A prevention trial is set up
with such children randomized to placebo or drug, and the asthma events
that developed over time are recorded in a diary. Typically, a~patient
has more than one asthma event. The different events are thus clustered
within a patient and ordered in time. This ordering can be taken into
account in the model. The data are presented in calendar time format,
where the time at risk for a particular event is the time from the end
of the previous event (asthma attack) to the start of the next event
(start of the next asthma attack). A particular patient has different
periods at risk during the total observation period which are separated
either by an asthmatic event that lasts one or more days or by a period
in which the patient was not under observation. The start and end of
each such risk period is required, together with the status indicator
to denote whether the end of the risk period corresponds to an asthma
attack or not.

%s3 ###
\section{Review of Key Ingredients}\label{ingredients}

In Section~\ref{glm} we will first describe the conventional
exponential family and generalized linear modeling based on it.
Section~\ref{overdispersion} is devoted to a brief review of models
for overdispersion. Section~\ref{normalre} focuses on the normally
distributed case.

%s3.1 ###
\subsection{Standard Generalized Linear Models}\label{glm}

A random variable $Y$ follows an exponential family distribution if the
density is of the form
%
%e1 ###
\begin{eqnarray}\label{mv03 density}
f(y) & \equiv& f(y | \eta, \phi)\nonumber
\\[-8pt]
\\[-8pt]
&=&   \exp \{ \phi
^{-1}[y\eta-\psi(\eta)]+c(y,\phi)  \}\nonumber
\end{eqnarray}
for a specific set of unknown parameters $\eta$ and $\phi$, and for
known functions $\psi(\cdot)$ and $c(\cdot,\cdot)$. Often, $\eta$
and $\phi$ are termed ``natural parameter'' (or ``canonical parameter'')
and ``dispersion parameter,'' respectively.

It can easily be shown (Molenberghs and Verbeke, \citeyear{Molenberghs2005}) that the first
two moments follow from the function $\psi(\cdot)$ as
%
%e3 ###
%e2 ###
\begin{eqnarray}\label{meanrel}
\mathrm{E}(Y)&=&\mu=\psi'(\eta),\\\label{varrel}
\operatorname{Var}(Y)&=&\sigma^2=\phi\psi''(\eta).
\end{eqnarray}
An important implication is that, in general, the mean and variance are
related through $\sigma^2 =\break  \phi\psi'' [\psi^{\prime-1} (\mu)] = \phi
v(\mu)$, with $v(\cdot)$ the so-called variance function, describing
the mean--variance relationship.

%t1 ###
\begin{sidewaystable*}
\tablewidth=\textwidth
\tabcolsep=0pt
\caption{Conventional exponential family members and extensions with conjugate random effects}
\label{grotetabel}
{\footnotesize
\begin{tabular*}{\textwidth}{@{\extracolsep{\fill}}lcccccc@{}}
\hline
\textbf{Element}&\textbf{Notation}&\textbf{Continuous}&\textbf{Binary}&\textbf{Count}&\multicolumn{2}{c@{}}{\textbf{Time to event}}\\
\hline
\multicolumn{7}{@{}l@{}}{Standard univariate exponential family}\\
\quad Model&&Normal&Bernoulli&Poisson&Exponential&Weibull\\
\quad Model&$f(y)$&
$\frac{1}{\sigma\sqrt{2\pi}}e^{- {(y-\mu)^2}/({2\sigma^2})}$&
$\pi^y (1-\pi)^{1-y}$&
$\frac{e^{-\lambda}\lambda^y}{y!}$&
$\varphi e^{-\varphi y }$&
$\varphi\rho y^{\rho-1}e^{-\varphi y^{\rho}}$\\
\quad Nat. param&$\eta$&
$\mu$&
$\ln [{\pi}/(1-\pi) ]$&
$\ln\lambda$&
$-\varphi$\\
\quad Mean function&$\psi(\eta)$&
$\eta^2/2$&
$ \ln[1 + \exp(\eta)]$&
$\lambda= \exp(\eta)$&
$-\ln(-\eta)$\\
\quad Norm. constant&$c(y,\phi)$&
$\frac{\ln(2 \pi\phi)}{2} - \frac{y^2}{2 \phi}$&
$0$&
$- \ln y!$&
$0$\\
\quad (Over)dispersion&$\phi$&
$\sigma^2$&
$1$&
$1$&
$1$
\\
\quad Mean&$\mu$&
$\mu$&
$\pi$&
$\lambda$&
$\varphi^{-1}$&
$\varphi^{-1/\rho}\Gamma(\rho^{-1}+1)$\\
\quad Variance&$\phi v(\mu)$&
$\sigma^2$&
$\pi(1-\pi)$&
$\lambda$&
$\varphi^{-2}$&
$\varphi^{-2/\rho} [\Gamma(2\rho^{-1}+1)-\Gamma(\rho
^{-1}+1)^2 ]$
\\[6pt]
\multicolumn{7}{@{}l@{}}{Exponential family with conjugate random effects}\\
\quad Model&&Normal--normal&Beta--binomial&Negative-binomial&Exponential--gamma&Weibull--gamma\\
\quad Hier. model&
$f(y|\theta)$&
$\frac{1}{\sigma\sqrt{2\pi}}e^{- {(y-\theta)^2}/({2\sigma^2})}$&
$\theta^y (1-\theta)^{1-y}$&
$\frac{e^{-\theta}\theta^y}{y!}$&
$\varphi\theta e^{-\varphi\theta y }$&
$\varphi\theta\rho y^{\rho-1}e^{-\varphi\theta y^{\rho}}$\\
\quad RE model&
$f(\theta)$&
$\frac{1}{\sqrt{d}\sqrt{2\pi}}e^{- {(\theta-\mu)^2}/({2d})}$&
$\frac{\theta^{\alpha-1}(1-\theta)^{\beta-1}}{B(\alpha,\beta)}$&
$\frac{\theta^{\alpha-1}e^{-\theta/\beta}}{\beta^\alpha\Gamma
(\alpha)}$&
$\frac{\theta^{\alpha-1}e^{-\theta/\beta}}{\beta^\alpha\Gamma
(\alpha)}$&
$\frac{\theta^{\alpha-1}e^{-\theta/\beta}}{\beta^\alpha\Gamma
(\alpha)}$\\
\quad Marg. model&
$f(y)$&
$\frac{1}{\sqrt{\sigma^2+d}\sqrt{2\pi}}e^{- {(y-\mu
)^2}/({2(\sigma^2+d)})}$&
$(\alpha+\beta)\frac{\Gamma(\alpha)}{\Gamma(\alpha+y)}\frac
{\Gamma(\beta)}{\Gamma(\beta+1-y)}$&
$\frac{\Gamma(\alpha+y)}{y!\Gamma(\alpha)}
 (
\frac{\beta}{\beta+1}
 )^{y}
 (
\frac{1}{\beta+1}
 )^{\alpha}$&
$\frac{
\varphi\alpha\beta
}
{
(1+\varphi\beta y)^{\alpha+1}
}
$
&$\frac{
\varphi\rho y^{\rho-1}\alpha\beta
}
{
(1+\varphi\beta y^\rho)^{\alpha+1}
}
$
\\
&$h(\theta)$&
$\theta$&
$\ln[\theta/(1-\theta)]$&
$\ln(\theta)$&
$-\theta$&
$-\theta$
\\
&$g(\theta)$&
$-\frac{1}{2}\theta^2$&
$-\ln(1-\theta)$&
$\theta$&
$-\ln(\theta)/\varphi$&
$-\ln(\theta)/\varphi$
\\
&$\phi$&
${\sigma^2}$&
$1$&
$1$&
$1/\varphi$&
$1/\varphi$
\\
&$\gamma$&
${1}/{d}$&
$\alpha+\beta-2$&
$1/\beta$&
$\varphi(\alpha-1)$&
$\varphi(\alpha-1)$
\\
&$\psi$&
$\mu$&
$\frac{\alpha-1}{\alpha+\beta-2}$&
$\beta(\alpha-1)$&
$[\beta\varphi(\alpha-1)]^{-1}$&
$[\beta\varphi(\alpha-1)]^{-1}$
\\
&$c(y,\phi)$&
$-\frac{1}{2}\phi y^2-\frac{1}{2}\ln (\frac{2\pi}{\phi
} )$&
$0$&
$-\ln(y!)$&
$\ln(\varphi)$&
$\ln (\varphi\rho y^{\rho-1} )$
\\
&$c^\ast(\gamma,\psi)$&
$-\frac{1}{2}\gamma\psi^2-\frac{1}{2}\ln (\frac{2\pi
}{\gamma} )$&
$-\ln B(\gamma\psi+1,$&
$(1+\gamma\psi)\ln\gamma$&
$\frac{\gamma+\varphi}{\varphi}\ln(\gamma\psi)-\ln\Gamma
(\frac{\gamma+\varphi}{\varphi} )$&
$\frac{\gamma+\varphi}{\varphi}\ln(\gamma\psi)-\ln\Gamma
(\frac{\gamma+\varphi}{\varphi} )$
\\
&&&\hspace*{40pt}$\gamma-\psi\gamma+1)$&\hspace*{10pt}\qquad$-\ln\Gamma(1+\gamma\psi)$\\
\quad Mean&$\rE(Y)$&
$\mu$&
$\frac{\alpha}{\alpha+\beta}$&
$\alpha\beta$&
$[\varphi(\alpha-1)\beta]^{-1}$&
$\frac{\Gamma(\alpha-\rho^{-1})\Gamma(\rho^{-1}+1)}{(\varphi
\beta)^{1/\rho}\Gamma(\alpha)}$
\\
\quad Variance&$\operatorname{Var}(Y)$&
$\sigma^2+d$&
$\frac{\alpha\beta}{(\alpha+\beta)^2}$&
$\alpha\beta(\beta+1)$&
$\alpha[\varphi^2(\alpha-1)^2(\alpha-2)\beta^2]^{-1}$&
$\frac{1}{\rho(\varphi\beta)^{2/\rho}\Gamma(\alpha)}
 [
2\Gamma(\alpha-2\rho^{-1})\Gamma(2\rho^{-1})
$
\\
&&&&&&
$ \hspace*{42pt}
 -\,
\frac{\Gamma(\alpha-\rho^{-1})^2\Gamma(\rho^{-1})^2}{\rho\Gamma
(\alpha)}
 ]$
\\
\hline
\end{tabular*}}
\end{sidewaystable*}

Key instances of the exponential family for normal, binary, count and
time-to-event data are listed in Table~\ref{grotetabel}, along with
their exponential family elements. The normal model is special, in
particular, also because the overdispersion parameter is needed to
allow for a variance other than unity. As a result, the mean--variance
relationship is absent for this model, but present for all others. In
the binary case, an alternative to the Bernoulli model with logit link
is the probit model, where $\eta=\Phi^{-1}(\pi)$ and $\Phi(\cdot)$
is the standard normal cumulative distribution function. Evidently,
this model is slightly less standard because the probit model is not
the natural link, as we will see in Section~\ref{binarycaselogit}, it
has appeal in the overdispersed and/or repeated contexts.

In the Weibull and exponential model, the decomposition $\varphi
=\lambda e^{\mu}$ is often employed, with notation as in Table~\ref
{grotetabel}, allowing for $\mu$ to become a function of covariates.
Evidently, here, while $\mu$ is a component of the mean function, it
is in itself not equal to the mean. Note also that the Weibull model
does not belong to the exponential family in a conventional sense,
unless in a somewhat contrived fashion where $y$ is replaced by $y^\rho
$. In the mean and variance expressions for the Weibull (Table~\ref
{grotetabel}), $\Gamma(\cdot)$ represents the gamma function.

In some situations, for example, when quasi-likeli\-hood methods are
employed (McCullagh and Nelder, \citeyear{McCullagh1989}; Molenberghs and Verbeke, \citeyear{Molenberghs2005}),
no full distributional assumptions are made, but one rather restricts
to specifying the first two moments (\ref{meanrel}) and (\ref
{varrel}). In such an instance, the variance function $v(\mu)$ can be
chosen in accordance with a particular member of the exponential
family. If not, then parameters cannot be estimated using maximum
likelihood principles. Instead, a set of estimating equations needs to
be specified, the solution of which is referred to as the
quasi-likelihood estimates.

In a regression context, where one wishes to explain variability
between outcome values based on measured covariate values, the model
needs to incorporate covariates. This leads to so-called generalized
linear models. Let $Y_1, \ldots, Y_N$ be a set of independent
outcomes, and let $\bm{x}_1, \ldots, \bm{x}_N$ represent the
corresponding $p$-dimensional vectors of covariate values. It is
assumed that all $Y_i$ have densities $f(y_i | \eta_i, \phi)$, which
belong to the exponential family, but a different natural parameter
$\eta_i$ is allowed per observation. Specification of the generalized
linear model is completed by modeling the means $\mu_i$ as functions
of the covariate values. More specifically, it is assumed that
$\mu_i= h(\eta_i)= h(\bm{x}_i'\bolds\xi)$,
for a known function $h(\cdot)$, and with $\bolds\xi$ a vector of $p$
fixed, unknown regression coefficients. Usually, $h^{-1}(\cdot)$ is
called the link function. In most applications, the so-called natural
link function is used, that is, $h(\cdot)= \psi'(\cdot)$, which is
equivalent to assuming $\eta_i=\bm{x}_i'\bolds\xi$. Hence, it is
assumed that the natural parameter satisfies a linear regression model.

%s3.2 ###
\subsection{Overdispersion Models}\label{overdispersion}

It is clear from Table~\ref{grotetabel} that the standard Bernoulli,
Poisson and exponential models force the mean and variance functions to
depend on a single parameter. However, comparing the sample average
with the sample variance might already reveal in certain applications
that this assumption is not in line with a particular set of data, for
count and time-to-event data, for example. While this is one of the
senses in which the binary case is somewhat exceptional, because a set
of i.i.d. Bernoulli data cannot contradict the mean--variance
relationship, it would still hold for the related binomial case, where
the data take the form of $n_i$ successes out of $z_i$ trials.

Therefore, a number of extensions have been proposed, as briefly
mentioned in the \hyperref[intro]{Introduction}. Hinde and Dem\'etrio (\citeyear{Hinde1998a,Hinde1998b}) provide
general treatments of overdispersion. The Poisson case received
particular attention by Breslow (\citeyear{Breslow1984}) and Lawless (\citeyear{Lawless1987}). Molenberghs
and Verbeke (\citeyear{Molenberghs2005}) mention various model-based approaches that
accommodate overdispersion, including the beta-binomial model (Skellam,
\citeyear{Skellam1948}), the Bahadur model (\citeyear{Bahadur1961}), the multivariate probit model (Dale,
\citeyear{Dale1986}; Molenberghs and Lesaffre, \citeyear{Molenberghs1994}) and certain versions of the
generalized linear mixed model (Breslow and Clayton, \citeyear{Breslow1993}). The latter
family will be studied in Section~\ref{normalre}.

A straightforward and commonly encountered step is to allow the
overdispersion parameter $\phi\ne1$, so that (\ref{varrel}) produces
$\operatorname{Var}(Y)=\phi v(\mu)$. This is in line with the moment-based
approach mentioned in the previous section, but can also be engendered
by fully parametric assumptions.

An elegant way forward is through a two-stage approach. For binary
data, one would assume that $Y_i|\pi_i\sim\operatorname{Bernoulli}(\pi_i)$
and further that $\pi_i$ is a random variable with $\mathrm{E}(\pi
_i)=\mu_i$ and $\operatorname{Var}(\pi_i)=\sigma^2_i$. Using iterated
expectations, it follows that
\begin{eqnarray*}
\rE(Y_i)
&=&\rE[\rE(Y_i|\pi_i)]=\rE(\pi_i)=\mu_i,\\
\Var(Y_i)
&=&\rE[\Var(Y_i|\pi_i)]+\Var[\rE(Y_i|\pi_i)]\\
&=&\rE[\pi_i(1-\pi_i)]+\Var(\pi_i)\\
&=&E(\pi_i)-E(\pi_i^2)+E(\pi_i^2)-E(\pi_i)^2\\
&=&\mu_i(1-\mu_i),
\end{eqnarray*}
underscoring that purely Bernoulli data are unable to capture overdispersion.

Likewise, for the Poisson case, we assume that $Y_i|\zeta_i\sim\mbox
{Poi}(\zeta_i)$ and then that $\zeta_i$ is a random variable with
$\mathrm{E}(\zeta_i)=\mu_i$ and $\operatorname{Var}(\zeta_i)=\sigma^2_i$.
Also here then, it follows that
\begin{eqnarray*}
\rE(Y_i)&=&\rE[\rE(Y_i|\zeta_i)]=\rE(\zeta_i)=\mu_i,\\
\Var(Y_i)&=&\rE[\Var(Y_i|\zeta_i)]+\Var[\rE(Y_i|\zeta_i)]\\
&=&\rE
(\zeta_i)+\Var(\zeta_i)=\mu_i+\sigma^2_i.
\end{eqnarray*}
Note that we have not assumed a particular distributional form for the
random effects $\pi_i$ and $\zeta_i$, respectively. Hence, this gives
rise to a semi-parametric specification. Similar routes can be followed
for other GLM, too.

In case it is considered advantageous to make full distributional
assumptions about the random effects, common choices are the beta
distribution for $\pi_i$ and the gamma distribution for $\zeta_i$; of
course, these are not the only ones.

Generally, the two-stage approach is made up of considering a
distribution for the outcome, given a random effect $f(y_i|\theta_i)$
which, combined with a model for the random effect, $f(\theta_i)$,
produces the marginal model:
%
%e4 ###
\begin{equation}
f(y_i)=\int f(y_i|\theta_i)f(\theta_i)\,d\theta_i.
\label{generalmarginalization}
\end{equation}

It is easy to extend this model to the case of repeated measurements.
We then assume a hierarchical data structure, where now $Y_{ij}$
denotes the $j$th outcome measured for cluster (subject) $i$, $i=1,
\ldots, N$, $j=1, \ldots, n_i$ and $\Yi$ is the $n_i$-dimensional
vector of all measurements available for cluster $i$. In the
repeated-measures case, the scalar $\zeta_i$ becomes a vector $\bfzeta
_i=(\zeta_{i1},\ldots,\zeta_{in_i})'$, with $\rE(\bfzeta_i)=\bfmu
_i$ and $\Var(\bfzeta_i)=\bfSigma_i$. For example, for the Poisson
case, similar logic as in the univariate case produces\break $\rE(\Yi
)=\bfmu_i$ and $\Var(\Yi)=M_i+\Sigma_i$, where $M_i$ is a diagonal
matrix with the vector $\bfmu_i$ along the diagonal. Note that a
diagonal structure of $M_i$ reflects the conditional independence
assumption: all dependence between measurements on the same unit stems
from the random effects. Generally, a versatile class of models
results. For example, assuming that the components of $\bfzeta_i$ are
independent, a pure overdispersion model follows, without correlation
between the repeated measures. On the other hand, assuming $\zeta
_{ij}=\zeta_i$, that is, that all components are equal, then $\Var
(\Yi)=M_i+\sigma^2_iJ_{n_i}$, where $J_{n_i}$ is an $(n_i\times n_i)$-dimensional matrix of ones. Such a structure can be seen as a general
version of compound symmetry. Of course, one can also combine general
correlation structures between the components of~$\bfzeta_i$.

Alternatively, this repeated version of the overdispersion model can be
combined with normal random effects in the linear predictor. This very
specific choice, proposed also by Thall and Vail (\citeyear{Thall1990}) and Dean
(\citeyear{Dean1991}), for the count case, will be the focus of the next section.

General marginalization (\ref{generalmarginalization}) may seem an
elegant and general principle, there is the issue of having to decide
which parameter to turn into a random one. This is especially true if
one considers the need to select an actual distributional form for the
random effect. A noteworthy exception is, as always, the linear mixed
model, combining a normal hierarchical model with a normal random
effect. It forms the basis of the two strands of random-effects models
that are potentially brought together in the combined models of
Section~\ref{combined}: on the one hand, normal random effects can be
considered with nonnormal outcomes, producing the GLMM; on the other
hand, gamma random effects for the Poisson model, beta random effects
with binomial data and gamma random effects for the Weibull model can
be considered. This is, seemingly, a disparate collection. However,
they are bound together by the property of \textit{conjugacy}, in the
sense of Cox and Hinkley (\citeyear{Cox1974}), page~370, and Lee, Nelder and
Pawitan (\citeyear{Lee2006}),\break page~178. The topic is also discussed by Agresti
(\citeyear{Agresti2002}). Informally, conjugacy refers to the fact that the hierarchical
and random-effects densities have similar algebraic forms. Conjugate
distributions produce a general and closed-form solution for the
corresponding marginal distribution.

We will first define conjugacy as is conventionally done, that is, in
models without the normal random effects and then, in Section~\ref
{combined}, introduce a further property, \textit{strong conjugacy},
necessary for situations where both normal and conventional conjugate
random effects are present. To simplify notation, we will provide the
definition at a general distribution level, with neither subject- nor
measurement-specific subscripts, so that it can be applied to both
univariate and longitudinal data. The hierarchical and random-effects
densities are said to be conjugate if and only if they can be written
in the generic forms
%
%e6 ###
%e5 ###
\begin{eqnarray}\label{conjugategenhier}\quad
f(y|\theta)&=&\exp \{\phi^{-1}[y h(\theta)-g(\theta
)]+c(y,\phi)  \},\\\label{conjugategenre}\quad
f(\theta) &=&\exp \{\gamma[\psi h(\theta)-g(\theta)]+c^\ast
(\gamma,\psi)  \},
\end{eqnarray}
where $g(\theta)$ and $h(\theta)$ are functions, $\phi$, $\gamma$
and $\psi$ are parameters, and the additional functions $c(y,\phi)$
and $c^\ast(\gamma,\psi)$ are so-called normalizing constants. It
can then be shown, upon constructing the joint distribution and then
integrating over the random effect, that the marginal model resulting
from (\ref{conjugategenhier}) and (\ref{conjugategenre}) equals
%
%e7 ###
\begin{eqnarray}\label{conjugategenmarg}
f(y)&=&\exp \biggl[c(y,\phi) + c^\ast(\gamma,\psi)\nonumber
\\[-8pt]
\\[-8pt]
&&\hphantom{\exp \biggl[} -c^\ast
\biggl(\phi^{-1}+\gamma,\frac{\phi^{-1} y+\gamma\psi}{\phi^{-1}+\gamma
} \biggr) \biggr].\nonumber
\end{eqnarray}

Table~\ref{grotetabel} gives model elements, such as density or
probability mass functions, conditional on random effects and
marginalized over these, as well as the random effects distributions.
For all models considered, the constants and functions featuring in
(\ref{conjugategenhier})--(\ref{conjugategenre}) are listed, and
finally marginal means and variances are provided. For some models,
these are well known (Hinde and Dem\'etrio, \citeyear{Hinde1998a,Hinde1998b}) and/or easy to
derive. For the time-to-event models, a sketch can be found in
Appendix~E. While there, the focus is on the combined
version of Section~\ref{combinedweibull}, the overdispersion case
considered here follows as a special case.

In the case of binary data, the model in Table~\ref{grotetabel} is the
familiar beta-binomial model. Note that the variance still obeys the
usual Bernoulli variance structure. This is entirely natural, given
that we still focus on a single binary outcome, in contrast to the more
conventional binomial basis model, where data of the format ``$z_i$
successes out $n_i$ trials'' are considered. We do not consider this
situation in this section, but rather leave it to Section~\ref
{combined}. In such a case, the variance structure becomes $\pi
_i(1-\pi_i)[1+\rho_i(n_i-1)]$, where $\rho_i$ is a measure for
correlation. All parameters, $p_i$ and $\rho_i$, can be expressed in
terms of $\alpha_i$ and $\beta_i$, ``cluster-specific'' versions of the
beta parameters.

For count data, the familiar negative-binomial\break model results. Unlike in
the binary case, univariate counts are able to violate the
mean--variance relationship of the Poisson distribution, hence the great
popularity of this and other types of models for overdispersion. The
same applies to the exponential distribution. Of course, already the
Weibull model, with its extra parameter $\rho$, alleviates the constraint.

The normal distribution case is a special one. Not only is it
self-conjugate, also the model is not identified, unlike all others.
This is because both random terms, seen from writing $Y_i=\mu
_i+b_i+\varepsilon_i$, are in direct, linear relationship with each
other. In the generalized linear context, the various random terms have
no direct linear alliance. The normal case will continue to be ``the odd
one out'' in models to come (Sections~\ref{normalre} and \ref{combined}).

The parameters $\alpha$ and $\beta$ in the beta and gamma
distributions are not always jointly identified. It is therefore
customary to impose restrictions, such as setting one of them equal to
a fixed value, for example, $\alpha=1$, or constraining their mean or
variance, etc. Such constraints operate differently, depending on other
elements present in the models. For example, the presence of additional
random effects in a model for repeated measures, such as in
Section~\ref{combined}, alters the meaning and restrictiveness of such
constraints.

Recall that the models at the bottom part of Table~\ref{grotetabel}
are not the only options, but rather common, elegant choices, where the
elegance draws to a large extent from conjugacy.

%s3.3 ###
\subsection{Models with Normal Random Effects}\label{normalre}

The generalized linear mixed model (Engel and Keen, \citeyear{Engel1994}; Breslow and
Clayton, \citeyear{Breslow1993}; Wolfinger and O'Connell, \citeyear{Wolfinger1993}) is likely the most
frequently used random-effects model in the context of perhaps
non-Gaussian repeated measurements. Not only is it a relatively
straightforward extension of the generalized linear model for
independent data (Section~\ref{glm}) to the context of hierarchically
organized data, on the one hand, and the linear mixed model (Verbeke
and Molenberghs, \citeyear{Verbeke2000}), on the other hand, but there is also a wide
range of software tools available for fitting such models.

Let $Y_{ij}$ be the $j$th outcome measured for cluster (subject) $i=1,
\ldots, N$, $j=1, \ldots, n_i$ and group the $n_i$ measurements into
a vector $\Y_i$. Assume that, in analogy with Section~\ref{glm},
conditionally upon $q$-dimensional random effects $\bi\sim N(\mathbf{0},
D)$, the outcomes $Y_{ij}$ are independent with densities of the form
%
%e8 ###
\begin{eqnarray}
\label{conddensity}
&&f_i(y_{ij} | \bi, \bolds\xi, \phi)\nonumber
\\[-8pt]
\\[-8pt]
 &&\quad = \exp \{ \phi
^{-1}[y_{ij}\lambda_{ij}-\psi(\lambda_{ij})]+c(y_{ij},\phi)
\},\nonumber
\end{eqnarray}
with
%
%e9 ###
\begin{eqnarray}
\label{linpred}
\eta[\psi'(\lambda_{ij})]&=&\eta(\mu_{ij})= \eta[E(Y_{ij} | \bi
,\bolds\xi)]\nonumber
\\[-8pt]
\\[-8pt]&=&\mathbf{x}_{ij}'\bolds\xi+ \mathbf{z}_{ij}'\bi\nonumber
\end{eqnarray}
for a known link function $\eta(\cdot)$, with $\mathbf{x}_{ij}$ and $\bm
{z}_{ij}$ $p$-dimensional and $q$-dimensional vectors of known
covariate values, with $\bolds\xi$ a $p$-dimensional vector of unknown
fixed regression coefficients, and with $\phi$ a scale
(overdispersion) parameter. Finally, let $f(\bi| D)$ be the density of
the $N(\mathbf{0}, D)$ distribution for the random effects $\bi$.

These models closely follow the ones formulated in the top part of
Table~\ref{grotetabel}, with key differences that now: (1) data
hierarchies are allowed for, in our setting owing to the longitudinal
collection of data; (2) the natural parameter is written as a linear
predictor, a function of both fixed and random effects.

Obviously, such models can be formulated for all data settings
considered in Table~\ref{grotetabel} and beyond. This is
conventionally done for continuous, Gaussian data, producing the linear
mixed-effects model (Verbeke and Molenberghs, \citeyear{Verbeke2000}), as well as for
binary data and counts. This kind of model is a bit less common for
survival data, where so-called frailty models (Duchateau and Janssen,
\citeyear{Duchateau2007}), rather of the type described in Section~\ref{overdispersion},
are more standard. Of course, also the accelerated failure time model
with random effects deserves mention, given that it takes the form of a
linear mixed model for logarithmic time.

We will not consider explicit expressions for such models here, because
they are relatively well studied (Fahrmeir and Tutz, \citeyear{Fahrmeir2001}; Molenberghs
and Verbeke, \citeyear{Molenberghs2005}) and, at any rate, conveniently follow as special
cases from the combined models of Section~\ref{combined}.

%s4 ###
\section{Models Combining Conjugate and Normal Random Effects}\label{combined}

%s4.1 ###
\subsection{General Model Formulation}\label{generalformulation}

Integrating both the overdispersion effects of Table~\ref{grotetabel}
(Section~\ref{overdispersion}) as well as the normal random effects of
Section~\ref{normalre} into the generalized linear model framework
produces the following general family:
%
%e10 ###
\begin{eqnarray}
\label{conddensitycombined}
&&f_i(y_{ij} | \bi, \bolds\xi,\theta_{ij},\phi)\nonumber
\\[-8pt]
\\[-8pt] &&\quad = \exp \{\phi
^{-1}[ y_{ij}\lambda_{ij}-\psi(\lambda_{ij})]+c(y_{ij},\phi)
\},\nonumber
\end{eqnarray}
with notation similar to the one used in (\ref{conddensity}), but now
with conditional mean
%
%e11 ###
\begin{equation}
\label{combinedgenmean}
E(Y_{ij}|\bi,\bolds\xi,\theta_{ij})=\mu^c_{ij}=\theta_{ij}\kappa_{ij},
\end{equation}
where the random variable $\theta_{ij}\sim{\cal G}_{ij}(\vartheta
_{ij},\sigma^2_{ij})$, $\kappa_{ij}=g(\mathbf{x}_{ij}'\bolds\xi+ \bm
{z}_{ij}'\bi)$, $\vartheta_{ij}$ is the mean of $\theta_{ij}$ and
$\sigma^2_{ij}$ is the corresponding variance. Finally, as before,
$\bi\sim N(\mathbf{0}, D)$. Write $\eta_{ij}=\mathbf{x}_{ij}'\bolds\xi+ \bm
{z}_{ij}'\bi$. Unlike in Section~\ref{normalre}, we now have two
different notations, $\eta_{ij}$ and $\lambda_{ij}$, to refer to the
linear predictor and/or the natural parameter. The reason is that
$\lambda_{ij}$ encompasses the random variables $\theta_{ij}$,
whereas $\eta_{ij}$ refers to the ``GLMM part'' only.

It is convenient, but not strictly necessary, to assume that the two
sets of random effects, $\bftheta_i$ and $\bi$, are independent of
each other. Regarding the components $\theta_{ij}$ of $\bftheta_i$,
three useful special cases result from assuming that: (1) they are
independent; (2) they are correlated, implying that the collection of
univariate distributions ${\cal G}_{ij}(\vartheta_{ij},\sigma
^2_{ij})$ needs to be replaced with a multivariate\vspace*{1pt} one; and (3) they
are equal to each other, useful in applications with exchangeable
outcomes $Y_{ij}$.

Obviously, parameterization (\ref{combinedgenmean}) allows for random
effects $\theta_{ij}$ capturing overdispersion, and formulated
directly at mean scale, such as described in Section~\ref
{overdispersion}, whereas $\kappa_{ij}$ could be considered the GLMM
component, as in Section~\ref{normalre}. The relationship between mean and
natural parameter now is
%
%e12 ###
\begin{equation}
\label{meannatpar}
\lambda_{ij}=h(\mu^c_{ij})=h(\theta_{ij}\kappa_{ij}).
\end{equation}
We can still apply standard GLM ideas, in particular, (\ref{meanrel})
and (\ref{varrel}), to derive the mean and variance, combined with
iterated-expectation-based calculations. For the mean, it follows that
%
%e13 ###
\begin{equation}
\label{meancombinedgeneral}
E(Y_{ij})=E(\theta_{ij})E(\kappa_{ij})=E[h^{-1}(\lambda_{ij})].
\end{equation}

%s4.2 ###
\subsection{Generic Approximations for Marginal Model Elements}\label
{genericapprox}

As we will see in ensuing specific cases (Sections~\ref
{normalcase}--\ref{combinedweibull}), (\ref{meancombinedgeneral})
allows for explicit expressions in a good number of cases. Generic
mean, variance and covariance approximations can be derived using the
expansion, around $\bi=\mathbf{0}$,
\[
\kappa_{ij}\approx g(\eta_{ij})+g'(\eta_{ij})\bz_{ij}'\bi+\tfrac
{1}{2}g''(\eta_{ij})\bz_{ij}'\bi\bi'\bz_{ij}.
\]
Details and expressions are provided in Appendix~A.

%s4.3 ###
\subsection{Strong Conjugacy}\label{strongconjugacy}

In Section~\ref{overdispersion} the concept of conjugacy was
introduced and exemplified in a number of cases (see\break Table~\ref
{grotetabel}). It is of interest to explore under what conditions
Model~(\ref{conddensitycombined}) still allows for conjugacy. The
complication is the presence of the multiplicative factor $\kappa
_{ij}$ in the mean structure. To make progress, we will study how
conjugacy plays out between\break Model~(\ref{conddensitycombined}) and the
distribution of the random effect $\theta_{ij}$, \textit{given} the
multiplicative factor $\kappa_{ij}$. In other words, conjugacy will be
considered conditional upon the normally-distributed random effect $\bi
$. To this effect, write (suppressing nonessential arguments\break from the functions)
%
%e14 ###
\begin{eqnarray}\label{conjugategenhiercombined}
f(y|\kappa\theta)&=&\exp \{\phi^{-1}[y h(\kappa\theta
)-g(\kappa\theta)]\nonumber
\\[-8pt]
\\[-8pt]&&\hspace*{74pt}{}+c(y,\phi)  \},\nonumber
\end{eqnarray}
generalizing (\ref{conjugategenhier}), and retain (\ref
{conjugategenre}). Applying the transformation theorem to (\ref
{conjugategenre}) leads to
\[
f(\theta|\gamma,\psi)=\kappa\cdot f(\kappa\theta|\widetilde
{\gamma},\widetilde{\psi}).
\]
Next, we request that the parametric form (\ref{conjugategenre}) be maintained:
%
%e15 ###
\begin{eqnarray}\label{conjugategenrewithk}
f(\kappa\theta) &=&\exp \{\gamma^\ast[\psi^\ast h(\kappa
\theta)-g(\kappa\theta)]\nonumber
\\[-8pt]
\\[-8pt]&&{}\hspace*{53pt}+c^{\ast\ast}(\gamma^\ast,\psi^\ast)
 \},\nonumber
\end{eqnarray}
where the parameters $\gamma^\ast$ and $\psi^\ast$ follow from
$\widetilde{\gamma}$ and $\widetilde{\psi}$ upon absorption of
$\kappa$. Then, the marginal model, in analogy with (\ref
{conjugategenmarg}), equals
%
%e16 ###
\begin{eqnarray}\label{conjugategenmargcombined}
f(y|\kappa)&=&\exp \biggl\{c(y,\phi) + c^{\ast\ast}(\gamma^\ast
,\psi^\ast)\nonumber
\\[-8pt]
\\[-8pt] &&\hphantom{\exp \biggl\{}{}+c^{\ast\ast} \biggl(\phi^{-1}+\gamma^\ast,\frac
{\phi^{-1} y+\gamma^\ast\psi^\ast}{\phi^{-1}+\gamma^\ast}
\biggr) \biggr\}.\nonumber
\end{eqnarray}
Evidently, not every model satisfying conjugacy in the sense of
Section~\ref{overdispersion} will allow for the present form of
conjugacy. We will refer to this condition as \textit{strong conjugacy}.
Examples include the normal, Poisson and Weibull (and hence
exponential) models with normal, gamma and gamma random effects,
respectively. A counterexample is provided by the Bernoulli, and hence
also binomial, model. Because the probit model does not allow for
conjugacy, not even in the usual sense, it is out of the picture here,
too. The latter does not preclude the existence of closed forms in the
probit case, as we will see in Section~\ref{binarycaseprobit}.

Note that the transition from strong conjugacy is a property entirely
of the random-effects distribution, and not of the data model, the
latter of which is needed, of course, for conjugacy itself.\vadjust{\goodbreak} For
example, for gamma random effects, we can write
%
%e17 ###
\begin{eqnarray}\label{strongconjgamma}
\frac{1}{\kappa} f(\theta|\alpha,\beta)&=&
\frac{1}{\kappa}\frac{1}{\beta^{\alpha}\Gamma(\alpha)}\theta
^{\alpha-1}e^{-\theta/\beta}\nonumber
\\&=&
\frac{1}{(\kappa\beta)^\alpha\Gamma(\alpha)}(\kappa\theta
)^{\alpha-1}e^{-(\kappa\theta)/(\kappa\beta)}\\
&=&f(\kappa\beta
|\alpha,\kappa\beta)\nonumber
\end{eqnarray}
and, hence, a scaled version of a gamma random effect is still a gamma
random effect, with retention of $\alpha$ and rescaling of $\beta$ to
$\kappa\beta$.

The importance of strong conjugacy lies, among others, in the easy
integration over the nonnormal random effects $\theta_{ij}$. As a
consequence, the resulting density is conditional on $\kappa$ and
hence on $\bi$ only, implying that standard software for generalized
linear or nonlinear mixed-effects models, such as the SAS procedure
NLMIXED, can be employed, a point to which we will return in
Section~\ref{estimation}.

We will now consider the normal, Poisson, binary and time-to-event
cases in turn. Details of the calculations for the Poisson case are
given in Molenberghs, Verbeke and Dem\'etrio (\citeyear{MolenberghsDem2007})
and summarized in
Appendix~B, while the binary and time-to-event cases
are supported by Appendices~C, D and E, respectively.

There is no need to spell out the various models in detail. The
different versions of (\ref{conddensitycombined}) follow
straightforwardly upon combining the models formulated in Table~\ref
{grotetabel} with the GLMM (\ref{conddensity}) and corresponding
linear predictor (\ref{linpred}). Precisely, the effect $\theta$
ought to be replaced by $\theta_{ij}\kappa_{ij}$, where $\kappa
_{ij}$ is defined by setting $\eta=\eta_{ij}$ equal to the linear
predictor whence $\kappa_{ij}$ is expressed, for the respective
models, as $\mu$, $\pi$, $\lambda$ and~$\phi$.

%s4.4 ###
\subsection{Specific Case: Continuous, Normally Distributed Data}\label
{normalcase}

The fully hierarchically specified linear mixed-ef\-fects model takes the
form (Verbeke and Molenberghs, \citeyear{Verbeke2000})
%
%e19 ###
%e18 ###
\begin{eqnarray}\label{lmmhier}
\BY_i|\bbi&\sim&N(X_i\bfbeta+Z_i\bbi,\Sigma_i),\\\label{lmmprior}
\bbi&\sim&N(0,D),
\end{eqnarray}
where $\bfbeta$ is a vector of fixed effects, and $X_i$ and $Z_i$ are
design matrices. The rows of $X_i\bfbeta+Z_i\bbi$ are made up by the
linear predictors (\ref{linpred}).

Based upon (\ref{lmmhier}) and (\ref{lmmprior}), the marginal model
can be derived:
%
%e20 ###
\begin{equation}\label{lmmmarg}
\BY_i\sim N(X_i\bfbeta,V_i=Z_i D Z_i'+\Sigma_i).
\end{equation}

We evidently consider a single set of random effects only, because, in
this case, the normal and conjugate random effects coincide, a unique
feature of the normal model. Strong conjugacy is a fortiori evident.

%s4.5 ###
\subsection{Specific Case: Poisson-Type Models for Count Data}\label
{poissoncase}

From the general developments above, the Poisson model with gamma and
normal random effects combined naturally follows. By way of overview,
let us assemble all model elements:
%
%e25 ###
%e24 ###
%e23 ###
%e22 ###
%e21 ###
\begin{eqnarray}\label{combined1}
Y_{ij}&\sim&\operatorname{Poi}(\theta_{ij}\kappa_{ij}),\\\label{combined2}
\kappa_{ij}&=&\exp (\bx_{ij}'\bfbeta+\bz_{ij}'\bi
),\\\label{combined3}
\bi&\sim&N(\mathbf{0},D),\\\label{combined4}
\rE(\bftheta_i)&=&\rE[(\theta_{i1},\ldots,\theta
_{in_i})']=\bfvartheta_i,\\\label{combined5}
\Var(\bftheta_i)&=&\Sigma_i.
\end{eqnarray}
This model has the same structure of the one by Booth et al.
(\citeyear{Booth2003}). In the spirit of Table~\ref{grotetabel}, the $\theta_{ij}$
can be assumed to follow a gamma model, producing, what we could term,
a Poisson--gamma--normal model or, equivalently, a
negative-binomial--normal model. When the gamma distribution is chosen,
it is implicitly assumed that the components $\theta_{ij}$ of
$\bftheta_i$ are independent. This is natural in many cases, in the
sense that the $\bi$ will induce association between repeated
measurements, with then the $\theta_{ij}$ taking care of additional
dispersion. In this case, $\Sigma_i$ reduces to a diagonal matrix.
Nevertheless, it is perfectly possible to allow for general covariance
structures. When a fully distributional specification would be desired,
then one could choose, for example, multivariate extensions of the
gamma model (Gentle, \citeyear{Gentle2003}).

As stated in general above, regarding the overdispersion random
effects, three situations could be of interest: (1) the random-effects
$\theta_{ij}$ are independent; (2) they are allowed to be dependent;
(3) they are equal to each other and hence reduce to $\theta
_{ij}\equiv\theta_i$.

The marginal mean vector and variance--covariance matrix are derived in
Appendix~B. The existence of such closed forms has
important implications because they allow, for example, for explicit
correlation expressions, on the one hand, and for a more versatile
collection of estimation methods, on the other hand, a point to which
we will return in Section~\ref{estimation}. The availability of
closed-form variance and joint-probability expressions supplements the
work of, for example, Zeger, Liang and Albert (\citeyear{Zeger1988}), who had stated
that only explicit mean expressions are available for a limited number
of generalized linear mixed models, other than the linear mixed model.

Let us consider strong conjugacy in this case. The corresponding model
elements in Table~\ref{grotetabel} change to
\begin{eqnarray*}
f(\theta)&=&\exp \biggl\{(\alpha-1)\ln\theta-\frac{1}{\beta
}\theta-\ln[\beta^{\alpha}\Gamma(\alpha)] \biggr\},\\
f(y|\lambda=\theta\kappa)&=&\exp \{
y\ln\theta-\kappa\theta-\ln y!+y\ln\kappa \},\\
\phi&=&1,\\
h(\theta)&=&\ln\theta,\\
g(\theta)&=&\theta\kappa,\\
\gamma&=&(\beta\kappa)^{-1},\\
\psi&=&\beta\kappa(\alpha-1),\\
c(y,\phi)&=&\ln y!+y\ln\kappa,\\
c^\ast(\gamma,\psi)&=&(1+\psi\gamma)\ln\gamma\kappa-\ln\Gamma
(1+\psi\gamma).
\end{eqnarray*}
Recall that the crux behind this result is (\ref{strongconjgamma}).

Even though Molenberghs, Verbeke and Dem\'etrio (\citeyear{MolenberghsDem2007}) did not do so,
it is fairly straightforward to derive the moments. Employing the
moments' expression for the standard Poisson (Johnson, Kemp and Kotz, \citeyear{Johnson2005},
page~162), the expression conditional upon the random effects is
%
%e26 ###
\begin{equation}
\label{momentstandardpoisson}
E(Y_{ij}^k)=\sum_{\ell=0}^kS(k,\ell)(\theta_{ij}\kappa_{ij})^\ell,
\end{equation}
where $S(k,\ell)$ is the so-called Stirling number of the second kind.
Integrating (\ref{momentstandardpoisson}) over the random effects
produces, without any problem,
%
%e27 ###
\begin{eqnarray}
E(Y_{ij}^k)&=&\sum_{\ell=0}^kS(k,\ell)\frac{\beta^\ell\Gamma
(\alpha+\ell)}{\Gamma(\alpha)}\nonumber
\\[-8pt]
\\[-8pt]
&&\hphantom{\sum_{\ell=0}^k}{}\cdot\exp \biggl[\ell\bx_{ij}'\bfbeta
+\frac{1}{2}\ell^2\bz_{ij}'D\bz_{ij} \biggr].\nonumber
\end{eqnarray}

%s4.6 ###
\subsection{Specific Case: Bernoulli-Type Models for Binary Data with
Logit Link}\label{binarycaselogit}

Similar to the Poisson case in Section~\ref{poissoncase}, a natural
binary-data counterpart to (\ref{combined1})--(\ref{combined5}) is
%
%e29 ###
%e28 ###
\begin{eqnarray}\label{bin1}
Y_{ij}&\sim&\operatorname{Bernoulli}(\pi_{ij}=\theta_{ij}\kappa
_{ij}),\\\label{bin2}
\kappa_{ij}&=&\frac{\exp (\bx_{ij}'\bfbeta+\bz_{ij}'\bi
 )}{
1+\exp (\bx_{ij}'\bfbeta+\bz_{ij}'\bi )},
\end{eqnarray}
completing the specification with (\ref{combined3})--(\ref{combined5}).
Unlike in the Poisson case, closed forms for neither the mean nor the
variance follow when normal random effects are present. When only
overdispersion random effects are included, especially when they are
assumed to follow a beta distribution, as in Table~\ref{grotetabel},
conjugacy applies. However, the beta distribution does not allow for
the multiplicative invariance as (\ref{strongconjgamma}), which will
preclude strong conjugacy.

When the overdispersion random effects are assumed to be equal, $\theta
_{ij}=\theta_i$, then the beta--binomial model would follow if no
normal random effects are present. The same is true, by the way, for
the\break compound-symmetry model generated by the hierarchical
random-intercepts model in the Gaussian case.

Explicitly considering $\theta_{ij}\sim\operatorname{Beta}(\alpha,\beta)$,
then $\phi_{ij}=\alpha/(\alpha+\beta)$, and
\begin{eqnarray*}
\sigma_{ij}^2&=&\sigma_{i,jj}=\frac{\alpha\beta}{(\alpha+\beta
)^2(\alpha+\beta+1)},\\
\sigma_{i,jk}&=&\rho_{ijk}\frac{\alpha\beta}{(\alpha+\beta
)^2(\alpha+\beta+1)}.
\end{eqnarray*}
Observe that there are two correlations: $\rho_{ijk}$, which described
the correlation between draws from the beta distribution and $(\alpha
+\beta+1)^{-1}$.
It is of course possible to let $\alpha$ and $\beta$ vary with $i$
and/or $j$. In such cases, the above and below expressions will change
somewhat, but computations are straightforward.

Using the general expressions, the above results can be used to derive
approximate expressions for means and variance--covariance elements. For
the special case of no normal random effects, but maintaining the fixed
effects in (\ref{bin2}), that is,
%
%e30 ###
\begin{equation}
\kappa_{ij}=\frac{\exp (\bx_{ij}'\bfbeta )}{
1+\exp (\bx_{ij}'\bfbeta )},
\label{binnore}
\end{equation}
we obtain
%
%e31 ###
\begin{eqnarray}\label{bernoullimean}
E(Y_{ij})&=&\frac{\alpha}{\alpha+\beta}\kappa_{ij},\nonumber\\
\operatorname{Var}(Y_{ij})&=&\frac{\alpha}{\alpha+\beta}\kappa_{ij}-
 \biggl(\frac{\alpha}{\alpha+\beta} \biggr)\kappa_{ij}^2,
\\
\operatorname{Cov}(Y_{ij},Y_{ik})&=&\rho_{ijk}\frac{\alpha\beta}{(\alpha
+\beta)^2(\alpha+\beta+1)}\kappa_{ij}\kappa_{ik}.\nonumber
\end{eqnarray}
If we further make exchangeability assumptions, that is, $\kappa
_{ij}=\kappa_{ik}\equiv\kappa_i$ and $\rho_{ijk}=\rho_i$, further
simplification follows. Finally, setting $\kappa_i=1$, the
conventional beta-binomial follows. It is then easy to derive the
resulting binomial version by defining
%
%e32 ###
\begin{equation}
\label{YtoZ}
Z_i=\sum_{i=1}^{n_i}Y_{ij}.
\end{equation}
Simple algebra then shows
\begin{eqnarray*}
E(Z_i)&=&n_i\frac{\alpha}{\alpha+\beta}=n_i\pi_i,\\
\operatorname{Var}(Z_i)&=&n_i\frac{\alpha\beta}{(\alpha+\beta)^2} \biggl\{
1+(n_i+1)\frac{1}{\alpha+\beta+1}
 \biggr\}\\
 &=&n_i\pi_i(1-\pi_i) \{1+(n_i-1)\widetilde{\rho
}_i \},
\end{eqnarray*}
with $\widetilde{\rho}_i$ the beta-binomial correlation. Hence, the
conventional beta-binomial model follows.

In comparison to the longitudinal Poisson case, the longitudinal binary
case appears to defeat closed-form solutions and strong conjugacy.
However, this hinges on the fact that we employ the logit link. In
spite of it being a very natural choice in the univariate case, it does
not combine very nicely with normal random effects. Recall that this is
known already from the GLMM framework for binary data. Therefore, it is
sensible to study the probit link instead. The random-effects probit
model has received some attention in earlier decades (Schall, \citeyear{Schall1991};
Guilkey and Murphy, \citeyear{Guilkey1993}; Hedeker and Gibbons, \citeyear{Hedeker1994}; McCulloch,
\citeyear{McCulloch1994};
Gibbons and Hedeker, \citeyear{Gibbons1997}; Renard, Molenberghs and Geys, \citeyear{Renard2004}), with
emphasis primarily on computational schemata to deal with the
multivariate normal integral. We will return to this aspect in
Section~\ref{estimation}.

%s4.7 ###
\subsection{Specific Case: Bernoulli-Type Models for Binary Data with
Probit Link}\label{binarycaseprobit}

Introducing the probit version of the model, while at the same time
assuming that the overdispersion parameters are beta distributed, comes
down to
%
%e34 ###
%e33 ###
\begin{eqnarray}\label
{probit2}
\kappa_{ij}&=&\Phi_1(\bx_{ij}'\bfbeta+\bz_{ij}'\bi),\\\label{probit3}
\theta_{ij}&\sim&\operatorname{Beta}(\alpha,\beta).
\end{eqnarray}
Like before, $\alpha$ and $\beta$ could be allowed to vary with $i$
and/or $j$.

It now follows that the joint distribution can be written as (details
in Appendix~D)
%
%e35 ###
\begin{equation}
\label{jointprobprobit}
f_{n_i}(\mathbf{y}_i={\bm1})= \biggl(\frac{\alpha}{\alpha
+\beta} \biggr)^{n_i}\cdot
\Phi_{n_i}(X_i\bfbeta;L_{n_i}^{-1}),
\end{equation}
with
%
%e36 ###
\begin{equation}
L_{n_i}=I_{n_i}-Z_i (D^{-1}+Z_i'Z_i )^{-1}Z_i'.
\label{lmatrix}
\end{equation}
More details on the cell probabilities, as well as on means and
variances, can be found in Appendix~D.

It is important to note that the existence of closed-form expressions
for the probit case opens a window of opportunity for the logit case.
Indeed, the well-known approximation formulae, linking the normal and
logistic densities, proves useful here. As shown in Johnson and Kotz
(\citeyear{Johnson1970}), page~6, and used in Zeger, Liang and Albert (\citeyear{Zeger1988}),
%
%e37 ###
\begin{equation}
\label{probitlogit}
\frac{e^y}{1+e^y}\approx\Phi_1(cy),
\end{equation}
with $c=(16\sqrt{3})/(15\pi)$.
Applied to (\ref{bin1})--(\ref{bin2}), we find
%
%e38 ###
\begin{eqnarray}\label{probitlogit2}
\pi_{ij}&\sim&\theta_{ij}
\frac{\exp (\bx_{ij}'\bfbeta+\bz_{ij}'\bi )}{
1+\exp (\bx_{ij}'\bfbeta+\bz_{ij}'\bi )}\nonumber
\\[-8pt]
\\[-8pt]
&\approx&
\theta_{ij}\Phi_1[c(\bx_{ij}'\bfbeta+\bz_{ij}'\bi)].
\nonumber
\end{eqnarray}
Applying (\ref{probitlogit2}) to (\ref{jointprobprobit}) yields
%
%e39 ###
\begin{equation}
\label{jointproblogit}
f_{n_i}(\mathbf{y}_i={\bm1})\approx \biggl(\frac{\alpha
}{\alpha+\beta} \biggr)^{n_i}\cdot
\Phi_{n_i} (cX_i\bfbeta;\widetilde{L}_{n_i}^{-1} ),
\end{equation}
with
\[
\widetilde{L}_{n_i}=I_{n_i}-c^2Z_i (D^{-1}+Z_i'Z_i )^{-1}Z_i'.
\]
For the expectation, we find, based on (\ref{probitlogit2}) and (D.4)
%e40 ###
\begin{eqnarray}\label{logitmean}
E(Y_{ij})&\approx&\frac{\alpha}{\alpha+\beta}\nonumber
\\[-8pt]
\\[-8pt]&&{}\cdot\Phi_1
(|I+c^2Dz_{ij}z_{ij}'|^{-1/2}c\bx_{ij}'\bfbeta ),\nonumber
\end{eqnarray}
with similar expressions for the variance and covariance terms.
Note that, upon estimating the parameters within the probit
approximation paradigm, back-transformation to the original logit scale
is possible, using expressions such as (\ref{probitlogit2}) and (\ref
{logitmean}). This opens perspectives for alternative estimation
methods for the combined model with logit link, with the important
special case of the normal-logistic GLMM.

In the Bernoulli case, calculating the moments is extremely simple.
Indeed, the Bernoulli moments are all identical. The conditional
moments are all $E(Y_{ij}^k|\theta_{ij},\bi)=\theta_{ij}\kappa
_{ij}$ ($k=1,2,\ldots$). Hence, they all reduce to (\ref
{bernoullimean}). In the probit case, they equal to (D.4).

%s4.8 ###
\subsection{Specific Case: Weibull- and Exponential-Type Models for
Time-to-Event Data}\label{combinedweibull}

The general Weibull model for repeated measures, with both gamma and
normal random effects, can be expressed as
%
%e43 ###
%e42 ###
%e41 ###
\begin{eqnarray}
\label{weibull1}
f(\mathbf{y}_i|\bftheta_i,\bi)&=&
\prod_{j=1}^{n_i}\lambda\rho\theta_{ij}y_{ij}^{\rho-1}e^{\bx
_{ij}'\bfbeta+\bz_{ij}'\bi}\nonumber
\\[-8pt]
\\[-8pt]
&&\hphantom{\prod_{j=1}^{n_i}}{}\cdot e^{-\lambda y_{ij}^{\rho}\theta_{ij}e^{\bx_{ij}'\bfbeta+\bz
_{ij}'\bi}},\nonumber\\
\label
{gengamma}\label{weibull2}
f(\bftheta_i)&=&\prod_{j=1}^{n_i}\frac{1}{\beta_j^{\alpha_j}\Gamma
(\alpha_j)}
\theta_{ij}^{\alpha_j-1}e^{-\theta_{ij}/\beta_j},\\\label{weibull3}
f(\bi)&=&\frac{1}{(2\pi)^{q/2}|D|^{1/2}}e^{-({1}/{2})\bi
'D^{-1}\bi}.
\end{eqnarray}
A few observations are in place. First, it is implicit that the gamma
random effects are independent. This need not be the case and, like in
the Poisson case, extension via multivariate gamma distributions is
possible. Second, setting $\rho=1$ leads to the special case of an
exponential time-to-event distribution. Third, it is evident that the
classical gamma frailty model (i.e., no normal random effects) and the
Weibull-based GLMM (i.e., no gamma random effects) follow as special
cases. Fourth, owing to the conjugacy result of Table~\ref{grotetabel}
and property~(\ref{strongconjgamma}) of the gamma density, strong
conjugacy applies. This is typically considered for the exponential
model, but it holds for the Weibull model too, merely by observing that
the Weibull model is nothing but an exponential model for the random
variable $Y_{ij}^\rho$. It is equally possible to derive this result
by merely rewriting the factor $\phi=\lambda\kappa$.
Fifth, the above expressions are derived for a two-parameter gamma
density. It is customary in a gamma frailty context (Duchateau and
Janssen, \citeyear{Duchateau2007}) to set $\alpha_j\beta_j=1$, for reasons of
identifiability. In this case, (\ref{gengamma}) is replaced by
%
%e44 ###
\begin{equation}
f(\bftheta_i)=\prod_{j=1}^{n_i}\frac{1}{ ( {1}/{\alpha
_j} )^{\alpha_j}\Gamma(\alpha_j)} \theta_{ij}^{\alpha
_j-1}e^{-\alpha_j\theta_{ij}}.\label{conventionalgamma}
\end{equation}
Alternatively, assuming $\alpha_j=1$ and $\beta_j=1/\delta_j$, one
could write
%
%e45 ###
\begin{equation}
f(\bftheta_i)=\prod_{j=1}^{n_i}\delta_je^{-\delta_j\theta_{ij}},
\label{alternativegamma}
\end{equation}
implying that the gamma density is reduced to an exponential one.
Closed-form expressions for the\break marginal density, means, variances,
covariances and moments are derived in Appendix~E,
where also a number of related facts are derived.

Of course, in this context of time-to-event data, further issues that
deserve attention are as follows: (1)~censoring and how to deal with
it; (2) derivation of related functions, such as the survivorship
function, as well as the hazard, cumulative hazard and intensity
functions; (3) the possibility of nonparametric baseline hazard
functions. These are nevertheless not considered here. While in
principle possible, we aim at focusing on commonality between various
GLM settings.

%s4.9 ###
\subsection{Implication for Computation of Correlation and Derived
Quantities}\label{derivedquantities}

Up to here, we have provided closed-form expressions for the marginal
joint distributions, the moments, and hence for means and variances,
for the normal, Poisson, probit and Weibull cases, with a combination
of normal random effects, on the one hand, supplemented, on the other
hand, with conjugate random effects, taking a normal, gamma, beta and
gamma form, respectively. The obvious one missing from the list is the
logit model, but then the logit-probit connection, as discussed in
Section~\ref{binarycaseprobit}, comes to rescue. Generally, progress
is possible whenever strong conjugacy applies.

These results and the ensuing calculations are useful for a number of
reasons, such as: (1) parameter estimation and derived inferences; (2)
implementation of estimation algorithms, as will be discussed in
Section~\ref{estimation}; and (3) the computation of derived quantities.

Such derived quantities include marginal correlation coefficients,
about which more detail is provided in the Appendix (Section~F). Of course, correlations are not always of direct
scientific interest and, when they are, one might not be willing to
base one's entire model choice on whether or not closed-form
correlations are available. That said, some considerations are in place.

First, our results indicate that closed-form correlations exist for a
number of commonly used models, such as the Poisson--normal GLMM and the
Weibull--gamma frailty model. Second, the same holds true for their
extensions within our proposed model. Third, when studying psychometric
reliability and generalizability (Vangeneugden et al., \citeyear{Vangeneugden2008a,Vangeneugden2010}),
the correlation function is the basic building block. Fourth,
correlation functions are also used in the context of surrogate marker
evaluation from clinical-trial data (Burzykowski, Molenberghs and
Buyse, \citeyear{Burzykowski2005}).

At the same time, the one important situation that evades direct
calculation of the marginal correlation is the logit with beta and
normal random effects, but then the probit--logit correspondence can be
invoked. On the one hand, the probit link can be used in lieu of the
logit link; on the other hand, the calculations can be carried out on
the probit scale, where after the results can be back transformed to
the logit scale.

Other key derived quantities include marginal regression parameters.
Suppose, for example, that one is interested in estimating the marginal
treatment effect from longitudinal clinical-trial data that are not
normally distributed. In principle, a marginal model could be fitted,
which oftentimes is done via generalized estimating equations (Liang
and Zeger, \citeyear{Liang1986}). However, when data are incomplete, such models pose
specific challenges even though remedies have been devised, such as
inverse probability weighting or a combination with multiple imputation
(for reviews, see Fitzmaurice et al., \citeyear{Fitzmaurice2009}). These, however,
come with their own problems. It is then attractive to fit a GLMM, with
or without additional random effects for overdispersion, and use the
closed-form mean expressions to derive marginal mean function. The
estimand of interest is then $E(\BY_i|T_i=1)-E(\BY_i|T_i=0)$, where
$T_i$ is the obvious indicator for the treatment to which the $i$th
subject has been assigned. Precision estimation then proceeds via the
delta method.

%s5 ###
\section{Estimation}\label{estimation}

A priori, fitting a combined model of the type described in
Section~\ref{combined} proceeds by integrating over the random
effects. The likelihood contribution of subject~$i$ is
%
%e46 ###
\begin{eqnarray}\label{lik contr}
&&f_i(\bm{y}_i| \bfvartheta,D,\bfvartheta_i,\Sigma_i)\nonumber\\
&&\quad  =
\int\prod
_{j=1}^{n_i}
 f_{ij}(y_{ij} | \bfvartheta,\bi,\bftheta_i)  f(\bi|
D)\\
&&\hphantom{\quad  =\int\prod
_{j=1}^{n_i}}{}\cdot f(\bftheta_i|\bfvartheta_i,\Sigma_i)  \,d \bi \,d\bftheta_i.\nonumber
\end{eqnarray}
Here, $\bfvartheta$ groups all parameters in the conditional model for
$\Y_i$. From (\ref{lik contr}) the likelihood derives as
%
%e47 ###
\begin{eqnarray}\label{likelihood GLMM}
&&L(\bfvartheta,D,\bfvartheta,\Sigma)\nonumber\\
 &&\quad  =  \prod_{i=1}^N
f_i(\bm{y}_i| \bfvartheta,D,\bfvartheta_i,\Sigma_i)
\nonumber
\\[-8pt]
\\[-8pt]
&& \quad =\prod_{i=1}^N \int\prod_{j=1}^{n_i}
f_{ij}(y_{ij} | \bfvartheta
,\bi,\bftheta_i)  f(\bi| D)\nonumber\\
&&\hphantom{\quad =\prod_{i=1}^N \int\prod_{j=1}^{n_i}}{}\cdot f(\bftheta_i|\bfvartheta_i,\Sigma
_i)  \,d \bi \,d\bftheta_i.\nonumber
\end{eqnarray}
The key problem in maximizing (\ref{likelihood GLMM}) is the presence
of $N$ integrals over the random effects $\bi$ and $\bftheta$. It is
widely claimed that the absence of a closed-form solution precludes an
analytical-integration based solution (Molenberghs and Verbeke, \citeyear{Molenberghs2005}),
explaining the popularity of Taylor-series expansion based methods,
such as PQL and MQL, Laplace approximation and numerical-integration
based methods. These have been implemented in, for example, the SAS
procedures GLIMMIX and NLMIXED. Several of the series expansion methods
tend to exhibit bias, an issue taken up in Breslow and Lin (\citeyear{Breslow1995}), and
suggesting the use of alternative methods.

However, thanks to our results in Section~\ref{combined}, further
progress can be made. Closed-form integration, apart from the normal
case, is within reach for the Poisson, probit and Weibull cases. Now,
some closed forms involve series expansions, and may be either time
consuming or cumbersome to implement. This notwithstanding, a variety
of alternative approaches are possible.

Let us turn to the Poisson case. While closed-form expressions can be
used to implement maximum likelihood estimation, with numerical
accuracy governed by the number of terms included in the series, one
can also proceed by what we will term partial marginalization. By this
we refer to integrating (\ref{combined1})--(\ref{combined5}) over the
gamma random effects only, leaving the normal random effects untouched.
The corresponding probability is
%
%e48 ###
\begin{eqnarray}
\label{probcombinedmargcond}
f(y_{ij}|\bi)&=&
\pmatrix{
\alpha_j+y_{ij}-1\cr \alpha_j-1
}
\cdot
 \biggl(
\frac
{
\beta_j
}
{
1+\kappa_{ij}\beta_j
}
 \biggr)^{y_{ij}}\nonumber
 \\[-8pt]
 \\[-8pt]
&&{}\cdot
 \biggl(
\frac
{
1
}
{
1+\kappa_{ij}\beta_{j}
}
 \biggr)^{\alpha_j}
\kappa_{ij}^{y_{ij}},\nonumber
\end{eqnarray}
where $\kappa_{ij}=\exp[\bx_{ij}'\bfbeta+\bz_{ij}'\bi]$. Note
that, with this approach, we assume that the gamma random effects are
independent within a subject. This is fine, given the correlation is
induced by the normal random effects.

Similarly, for the Weibull case we obtain
%
%e49 ###
\begin{equation}
\label{probcombinedmargcondweibull}
f(y_{ij}|\bi)=\frac
{
\lambda\kappa_{ij}e^{\mu_{ij}}\rho y_{ij}^{\rho-1}\alpha_j\beta_j
}
{
(1+\lambda\kappa_{ij}e^{\mu_{ij}}\beta_j y_{ij}^\rho)^{\alpha_j+1}
}.
\end{equation}
Because there is lack of strong conjugacy, the logit case defies the
mere exploitation of conjugate form, such as the negative-binomial form
(\ref{probcombinedmargcond}) and the Weibull--gamma frailty form (\ref
{probcombinedmargcondweibull}). Nevertheless, it is easy to derive, for
this case,
%
%e50 ###
\begin{eqnarray}
\label{probcombinedmargcondlogit}
f(y_{ij}|\bi)&=&\frac{1}{\alpha_j+\beta_j}\cdot(\kappa_{ij}\alpha
_j)^{y_{ij}}\nonumber
\\[-8pt]
\\[-8pt]
&&{}\cdot
[(1-\kappa_{ij})\alpha_j+\beta_j]^{1-y_{ij}}.\nonumber
\end{eqnarray}

For all of these, it is straightforward to obtain the fully
marginalized probability by numerically integrating the normal random
effects out of (\ref{probcombinedmargcond}),
(\ref{probcombinedmargcondweibull}) and
(\ref{probcombinedmargcondlogit}), using a tool such as the SAS
procedure NLMIXED that allows for normal random effects in arbitrary,
user-specified models.

The concept of partial integration always applies whenever strong
conjugacy holds. Indeed, an expression of the form (\ref
{conjugategenmargcombined}) corresponds to integrating over the
conjugate random effect $\theta$, while leaving the normally
distributed random effect embedded in the predictor, $\kappa$ in this
notation. Recall that, while expressions of the type (\ref
{conjugategenmargcombined}) appear to be for the univariate case, they
extend without problem to the longitudinal setting as well.

For the specific case of the marginalized probit model, the
computational challenge stems from the presence of a multivariate
normal integral of the form~(\ref{jointprobprobit}), a phenomenon also
known from the fully marginally specified multivariate probit model
(Ashford and Sowden, \citeyear{Ashford1970}; Lesaffre and Molenberghs, \citeyear{Lesaffre1991}; Molenberghs
and Verbeke, \citeyear{Molenberghs2005}). Specific to the context of the probit models with
random effects, Zeger, Liang and Albert (\citeyear{Zeger1988}) derived the marginal
mean function, needed for their application of generalized estimating
equations as a fitting algorithm for the marginalized probit model. It
is one of the first instances of the use of GEE to a nonmarginally
specified model. Precisely, these authors derive the marginal mean
function and (a working version of) the marginal variance--covariance
matrix. These are sufficient to implement GEE or, with appropriate
extension, also second-order GEE. Note that our derivations yield, for
strong conjugate cases in general, as well as for a number of
particular cases, not only the marginal mean and variance, but also all
moments and the entire joint distribution. Evidently, this is plenty to
implement GEE, but the other methods, described in this section, come
within reach, too.

In the same spirit, pseudo-likelihood can be used (Aerts et al., \citeyear{Aerts2002};
Molenberghs and Verbeke, \citeyear{Molenberghs2005}). This is particularly useful
when the joint marginal distribution is available but cumbersome to
manipulate and evaluate, such as in the probit case. This is the idea
followed by Renard, Molenberghs and Geys (\citeyear{Renard2004}) for a multilevel probit
model with random effects, similar in spirit to the probit models
considered in Section~\ref{binarycaseprobit}. Essentially, the joint
distribution is replaced with a product of factors of marginal and/or
conditional distributions of lower dimensions. Because such a product
does not necessarily recompose the original joint distribution,
sandwich-estimator ideas are then used to provide not only valid point
estimates, but also precision estimates and inferences derived therefrom.

Schall (\citeyear{Schall1991}) proposed an efficient and general estimation algorithm,
based on Harville's (\citeyear{Harville1974}) modification of Henderson's (\citeyear{Henderson1984})
mixed-model equations. Hedeker and Gibbons (\citeyear{Hedeker1994}) and Gibbons and
Hedeker (\citeyear{Gibbons1997}) proposed numerical-integration based methods, thus
considering neither marginal moments (means, variances) nor
marginalized joint probabilities. Guilkey and Murphy (\citeyear{Guilkey1993}) provide a
useful early overview of estimation methods and then revert to Butler
and Moffit's (\citeyear{Butler1982}) Hermite-integration based method, supplemented with
Monte Carlo Markov Chain ideas.

Further, one might, for example, opt for fully Bayes\-ian inferences.
Alternatively, the EM algorithm can be used, in line with Booth et al. (\citeyear{Booth2003}) for the Poisson case. The EM is a flexible framework
within which either the conjugate, or the normal, or both sets of
random effects can be considered the ``missing'' data over which
expectations are taken.

Booth et al. (\citeyear{Booth2003}) also considered nonparametric maximum
likelihood, in the spirit of Aitkin (\citeyear{Aitkin1999}) and Alf\`o and Aitkin
(\citeyear{Alf2000}). In addition, ideas of hierarchical generalized linear models
(Lee and Nelder, \citeyear{Lee1996,Lee2001a,Lee2001b,Lee2003}; Yun, Sohn and Lee, \citeyear{Yun2006};
Lee, Nelder and Pawitan, \citeyear{Lee2006}) can be employed.

A suite of methods is available that employ transformation results,
essentially based on transforming the nonnormal random effects to
normal ones, or vice versa.
To briefly describe these, write the contribution for subject $i$ to
the likelihood as
%
%e51 ###
\begin{equation}
L_i=\int \biggl[\prod_j f(y_{ij}|u_i) \biggr]p_u(u_i)\,du_i,
\label{drie}
\end{equation}
where $f(\cdot)$ specifies the outcome model given the random effects.
Furthermore, $p_u(\cdot)$ denotes the density of the random effect,
typically nonnormal. While the latter random effect can be
vector-valued, let us illustrate the method for the scalar case. To
simplify notation further, in (\ref{drie}), covariates and parameter
vectors have been suppressed from notation. Liu and Yu (\citeyear{Liu2008}) advocate
a simple transformation:
%
%e52 ###
\begin{equation}
L_i=\int \biggl[\prod_j f(y_{ij}|a_i) \biggr]\frac{p_u(a_i)}{\phi
(a_i)}\phi(a_i)\,da_i,
\label{vier}
\end{equation}
where now $a_i$ is a normal random effect. Evidently, $\phi(\cdot)$
is the standard (multivariate or univariate) normal density.
Liu and Yu (\citeyear{Liu2008}) complete their argument by stating that then the new model
\[
\biggl [\prod_j f(y_{ij}|a_i) \biggr]\frac{p_u(a_i)}{\phi(a_i)}
\]
can be subjected to the conventional quadrature techniques available
in, for example, SAS' NLMIXED procedure. A number of SAS
implementations for important particular cases are offered by these
authors. Obviously, the method can be expanded to our situation, where
apart from the nonnormal random effects, also normal random effects
are present. The justification of the method simply follows by applying
the transformation theorem at the level of the densities involved. The
usefulness of this method cannot be overestimated. It is especially
useful when partial integration is not possible, for example, when
strong conjugacy does not hold, like in the binary beta--normal--logit case.

Alternatively, Nelson et al. (\citeyear{Nelson2006}) advocate the transformation
%
%e53 ###
\begin{equation}
u_i=F_u^{-1}[(\Phi(a_i)],
\label{twee}
\end{equation}
where $F_u$ is the cumulative distribution function (CDF) of $u_i$ and
$\Phi(\cdot)$ is the standard normal CDF, as before. Nelson  et
al.'s method, labeled \textit{probability integral transformation}
(PIT), comes down to generating normal variates and then inserting
these in the model only after transformation (\ref{twee}), ensuring
that they are of the desired nature. It is tautologically clear that
(\ref{twee}) automatically ensures the support of the variable is
correctly mapped along with the variable itself. By passing through the
unit interval, by means of $\Phi(\cdot)$, and then applying
$F_u(\cdot)$, one forces, for example, a gamma variable to range over
the positive half line, a beta variable to be confined to the unit
interval, etc., as it should.\looseness=1

Lin and Lee (\citeyear{Lin2008}) present estimation methods for the specific case of
linear mixed models with skew-normal, rather than normal, random effects.

Quite apart from the choice of estimation method, it is important to
realize that not all parameters may be simultaneously identifiable. For
example, the gamma-distribution parameters in the Poisson case, $\alpha
$ and $\beta$, are not simultaneously identifiable when the
linear-predictor part is also present, because there is aliasing with
the intercept term. Therefore, one can set, for example, $\beta$ equal
to a constant, removing the identifiability problem. It is then clear
that $\alpha$, in the univariate case, or the set of $\alpha_j$ in
the repeated-measures case, describe the additional overdispersion, in
addition to what stems from the normal random effect(s). A similar
phenomenon also plays in the binary case, where both beta-distribution
parameters are not simultaneously estimable.

%s6 ###
\section{Analysis of Case Studies}\label{analysiscasestudies}

%s6.1 ###
\subsection{A Clinical Trial in Epileptic Patients}\label{analysisepilepsy}

We will analyze the epilepsy data, introduced in Section~\ref
{dataepilepsy}. Note that the data were analyzed before in Molenberghs
and Verbeke (\citeyear{Molenberghs2005}), Chapter~19, using generalized estimating equations
(Liang and Zeger, \citeyear{Liang1986}) and the Poisson--normal model. These authors used
a slightly different parameterization.

Let $Y_{ij}$ represent the number of epileptic seizures patient $i$
experiences during week $j$ of the follow-up period. Also, let $t_{ij}$
be the time-point at which $Y_{ij}$ has been measured, $t_{ij}= 1, 2,
\ldots,$ until at most $27$. Let us consider the combined model (\ref
{combined1})--(\ref{combined5}), with specific choices
%
%e54 ###
\begin{eqnarray} \label{mv19 glmm model}\quad
\ln(\kappa_{ij}) & = &  \cases{ (\xi_{00} + b_{i}) + \xi_{01} t_{ij}, &  if
placebo,\cr
(\xi_{10} + b_{i}) + \xi_{11}t_{ij}, &  if treated,
}
\end{eqnarray}
where the random intercept $b_i$ is assumed to be zero-mean normally
distributed with variance $d$.
We consider special cases: (1) the ordinary Poisson model, (2)~the
negative-binomial model, (3) the Poisson--normal model, together with
(4) the combined model. Estimates (standard errors) are presented in
Table~\ref{epilepsy estimates}. Clearly, both the negative-binomial
model and the Poisson--normal model are important improvements, in terms
of the likelihood, relative to the ordinary Poisson model. This should
come as no surprise since the latter unrealistically assumes there is
neither overdispersion nor correlation within the outcomes, while
clearly both are present. In addition, when considering the combined
model, there is a very strong improvement in fit when gamma and normal
random \mbox{effects} are simultaneously allowed for. This strongly affects
the point and precision estimates of such key parameters as the slope
difference and the slope ratio. There is also an impact on hypothesis
testing. The Poisson model leads to unequivocal significance for both
the difference ($p=0.0008$) and ratio ($p=0.0038$), whereas for the
Poisson--normal this is not the case for the difference of the slopes
($p=0.7115$), while some significance is maintained for the ratio
($p=0.0376$). Because the Poisson--normal is commonly used, it is likely
that in practice one would decide in favor of a treatment effect when
considering the slope ratio. This is no longer true with the
negative-binomial model, where the $p$-values change to $p=0.01310$ and
$p=0.2815$, respectively. Of course, one must not forget that, while
the negative-binomial model accommodates overdispersion, the $\theta
_{ij}$ random effects are assumed independent, implying independence
between repeated measures. Again, this is not realistic and, therefore,
the combined model is a more viable candidate, corroborated further by
the aforementioned likelihood comparison. This model produces
nonsignificant $p$-values of $p=0.2260$ and $p=0.1591$, respectively.%\looseness=1

%
%t2 ###
\begin{table*}
\tabcolsep=0pt
\tablewidth=410pt
\caption{Epilepsy study. Parameter estimates and
standard errors for the regression coefficients in (1) the Poisson
model, (2) the negative-binomial model, (3) the Poisson--normal model
and (4) the combined model. Estimation was done~by maximum likelihood
using numerical integration over the normal random effect, if
present}\label{epilepsy estimates}
\begin{tabular*}{410pt}{@{\extracolsep{\fill}}lcd{2.12}d{2.12}@{}}
\hline
&&\multicolumn{2}{c@{}}{\textbf{Estimate (s.e.)}}\\
\cline{3-4}
\textbf{Effect} & \textbf{Parameter}  & \multicolumn{1}{c}{\textbf{Poisson}} &
\multicolumn{1}{c@{}}{\textbf{Negative-binomial}}\\
\hline
Intercept placebo & $\xi_{00}$ & 1.2662\ (0.0424) & 1.2594\ (0.1119) \\
Slope placebo & $\xi_{01}$ & -0.0134\ (0.0043) & -0.0126\ (0.0111) \\
Intercept treatment & $\xi_{10}$ & 1.4531\ (0.0383) & 1.4750\ (0.1093) \\
Slope treatment & $\xi_{11}$ & -0.0328\ (0.0038) & -0.0352\ (0.0101)
\\
Negative-binomial parameter & $\alpha_1$ & \multicolumn{1}{c}{---} &
0.5274\ (0.0255) \\
Negative-binomial parameter & $\alpha_2=1/\alpha_1$ & \multicolumn
{1}{c}{---} & 1.8961\ (0.0918) \\
$-$2log-likelihood &&\multicolumn{1}{c}{$-$1492}&\multicolumn
{1}{c@{}}{$-$6755}\\[6pt]
& & \multicolumn{1}{c}{Poisson--normal} & \multicolumn
{1}{c@{}}{Combined}\\[6pt]
%Effect & Parameter & Estimate (s.e.)  & Estimate (s.e.)   \\
Intercept placebo & $\xi_0$ & 0.8179\ (0.1677) &0.9112\ (0.1755) \\
Slope placebo & $\xi_1$ & -0.0143\ (0.0044) & -0.0248\ (0.0077) \\
Intercept treatment & $\xi_0$ & 0.6475\ (0.1701) &0.6555\ (0.1782) \\
Slope treatment & $\xi_2$ & -0.0120\ (0.0043) & -0.0118\ (0.0074) \\
Negative-binomial parameter & $\alpha_1$ &\multicolumn{1}{c}{---} &
2.4640\ (0.2113) \\
Negative-binomial parameter & $\alpha_2=1/\alpha_1$ &\multicolumn
{1}{c}{---} & 0.4059\ (0.0348) \\
Variance of random intercepts & $d$ & 1.1568\ (0.1844) & 1.1289\ (0.1850)
\\
$-$2log-likelihood &&\multicolumn{1}{c}{$-$6810}&\multicolumn
{1}{c@{}}{$-$7664}\\
\hline
\end{tabular*}
\vspace*{-3pt}
\end{table*}

Thus, in conclusion, whereas the conventionally used and broadly
implemented Poisson--normal model would suggest a significant effect of
treatment, our combined model issues a message of caution, because
there is no evidence whatsoever regarding a treatment difference.

Molenberghs and Verbeke (\citeyear{Molenberghs2005}), Chapter~19, considered a Poisson--normal
model with random intercepts as well as random slopes in time. It is
interesting to note that, when allowing for such an extension in our
models, the random slopes improve the fit of the Poisson--normal model
with random intercept, but not of the combined one with random
intercept (details not shown).
As a consequence, the combined model with random intercept is the best
fitting one. At the same time, note that fitting such a model
establishes that the presence of a conjugate random effect does not
preclude the consideration of normal random effects beyond random intercepts.

Recall that the data were analyzed, too, by Booth et al.
(\citeyear{Booth2003}). While we considered four different models, these authors
focused on the Poisson--normal and combined implementations. There are
further differences in actual fixed-effects and random-effects models
considered, as well as in us further considering inferences for
differences and ratios.

%t3 ###
\begin{table*}
\tabcolsep=0pt
\tablewidth=355pt
\caption{Epilepsy study. Observed smallest and largest
values for the correlation function, for the Poisson--normal and
combined models, and for both treatment arms. The time pair for
which the values are observed is shown too (RI---random intercept; RS---random slope)}
\label{corrlimits}
\begin{tabular*}{355pt}{@{\extracolsep{\fill}}lccccc@{}}
\hline
& & \multicolumn{2}{c}{\textbf{Smallest value}} & \multicolumn{2}{c@{}}{\textbf{Largest
value}}\\
\ccline{3-4,5-6}
\textbf{Model}&\textbf{Arm}&$\bolds\rho$&\textbf{Time pair}&$\bolds\rho$&\textbf{Time pair}\\
\hline
Poisson--normal, RI &Placebo &0.8577&26 \& 27&0.8960& 1 \& 2\\
Poisson--normal, RI &Treatment&0.8438&26 \& 27&0.8794& 1 \& 2\\[6pt]
Combined, RI &Placebo &0.8259&26 \& 27&0.8981& 1 \& 2\\
Combined, RI &Treatment&0.8383&26 \& 27&0.8744& 1 \& 2\\[6pt]
Poisson--normal, RI$+$RS&Placebo &0.2966& \phantom{2}1 \& 27&0.9512& 26 \& 27\\
Poisson--normal, RI$+$RS&Treatment&0.2936& \phantom{2}1 \& 27&0.9530& 26 \& 27\\[6pt]
Combined, RI$+$RS &Placebo &0.4268& \phantom{2}1 \& 27&0.9281& 26 \& 27\\
Combined, RI$+$RS &Treatment&0.4225& \phantom{2}1 \& 27&0.9329& 26 \& 27\\
\hline
\end{tabular*}
\vspace*{-3pt}
\end{table*}

Let us now turn to the correlation functions. Given that the gamma
random effects are assumed independent, we only need to consider the
Poisson--normal and combined cases; the versions with and without random
slopes are considered. Obviously, because the fixed-effects structure
is not constant but rather depends on time, we have to apply the
general correlation function~(F.13). In the Poisson--normal
case with random intercepts only, and for the placebo group, based on
the parameter estimates in Table~\ref{epilepsy estimates}, we obtain
\begin{eqnarray*}
\Corr(Y(t),Y(s))&=&
{
35.58\cdot0.99^{t+s}
}\\
&&{}\big/
\bigl(\sqrt
{
(4.04\cdot0.99^t+35.58\cdot0.97^t)}\\&&{}\cdot\sqrt{
(4.04\cdot0.99^s+35.58\cdot0.97^s)
}\bigr)
,
\end{eqnarray*}
where $Y(t)$ represents the outcome for an arbitrary subject at time
$t$. Calculations in all other cases are similar. The smallest and
largest values for the correlation functions, for both arms, for both
the Poisson--normal and combined models, and for both choices of the
random-effects structure are given in Table~\ref{corrlimits}. When
only random intercepts are considered, the correlations range over a
narrow interval; they are rather high and there is little difference
between the Poisson--normal and combined models. However, turning to the
models with random intercepts and random slopes, several differences
become apparent. First, the values exhibit a much broader range between
their smallest and largest values. Second, the range is somewhat
overestimated by the Poisson--normal model, which then narrows when we
switch to the combined model, thereby incorporating overdispersion
effects, random intercepts and random slopes. Thus, the random slope
allows for the correlation to range over a considerable interval, while
the overdispersion effect avoids the range to be overly wide.

Within each model, there is relatively little difference between the
placebo and treated groups, although the difference is a bit more
pronounced in the combined model. Further, the correlation range within
every group is relatively narrow. The most noteworthy feature,
unquestionably, is the large discrepancy between both models. This is
because the Poisson--normal model forces the correlation and\break
overdispersion effects to stem from a single additional parameter, the
random-intercept variance~$d$. Thus, considerable overdispersion also
forces the correlation to increase, arguably beyond what is consistent
with the data. In the combined model, in contrast, there are \textit{two} additional parameters, giving proper justice to both correlation and
overdispersion effects. It was already clear from the above discussion
and that in Molenberghs, Verbeke and Dem\'etrio (\citeyear{MolenberghsDem2007}) that the
combined model is an important improvement. This now clearly manifests
itself in the correlation function, too.

%s6.2 ###
\subsection{A Clinical Trial in Onychomycosis}\label{analysisonychomycosis}

We will analyze the binary onychomycosis data, introduced in
Section~\ref{dataonychomycosis}.
For the logit, consider the model
%
%e55 ###
\begin{eqnarray}
Y_{ij} | (b_i) & \sim& \operatorname{Bernoulli}(\pi_{ij}), \nonumber\\\label{toenail mixed}
\operatorname{logit}(\pi_{ij})
& = & \xi_1 (1-T_i) + b_{i} + \xi_2 (1-T_i)
t_{ij}\nonumber
\\[-8pt]
\\[-8pt]&&{} + \xi_3 T_i + \xi_4 T_i t_{ij},\nonumber
\end{eqnarray}
where $T_i$ is the treatment indicator for subject $i$, $t_{ij}$ is the
time-point at which the $j$th measurement is taken for the $i$th
subject, and $b_i\sim N(0,d)$. Parameter estimates for the logistic
model, with and without the normal random effect, on the one hand, and
with and without the beta--binomial component, on the other hand, as
described in Section~\ref{binarycaselogit}, are presented in
Table~\ref{toenail estimates}. Observe that the model becomes hard to
fit when the beta random effects are present, which is seen from
estimates and standard errors in both the beta--binomial model as well
as the combined model. To understand this, we must observe that the
conjugate random effects in the Bernoulli case, unlike in the Poisson,
binomial and Weibull cases, cannot add to the variability, only to the
correlation structure. This means that there is considerably less
information available than in the other cases. This does not mean that
the beta random effects are unnecessary, but rather that they challenge
the stable estimation of other model parameters.
%
%t4 ###
\begin{table*}
\tabcolsep=0pt
\tablewidth=400pt
\caption{Onychomycosis study. Parameter estimates and
standard errors for the regression coefficients in (1) the logistic
model, (2) the beta--binomial model, (3) the logistic--normal model and
(4) the combined model. Estimation was done by maximum likelihood using
numerical integration over the normal random effect, if present}\label
{toenail estimates}
\begin{tabular*}{400pt}{@{\extracolsep{\fill}}lcd{2.12}d{3.13}@{}}
\hline
& &\multicolumn{2}{c@{}}{\textbf{Estimate (s.e.)}}\\
\cline{3-4}
\textbf{Effect} & \textbf{Parameter} &  \multicolumn{1}{c}{\textbf{Logistic}} & \multicolumn{1}{c@{}}{\textbf{Beta--binomial}}\\
\hline
Intercept treatment A & $\xi_{0}$ & -0.5571\ (0.1090) & 17.9714\
(1482.6) \\
Slope treatment A & $\xi_{1}$ & -0.1769\ (0.0246) & 5.2454\ (12970.0)
\\
Intercept treatment B & $\xi_{2}$ & -0.5335\ (0.1122) & 18.6744\
(2077.13) \\
Slope treatment B & $\xi_{3}$ & -0.2549\ (0.0309) & 4.7775\ (12912.0)
\\
Std. dev. random effect & $\sqrt{d}$ & \multicolumn
{1}{c}{---}&\multicolumn{1}{c}{---} \\
Ratio & $\alpha/\beta$ &\multicolumn{1}{c}{---} & 3.6739\ (0.2051) \\
$-$2log-likelihood &&\multicolumn{1}{c@{}}{1812}&\multicolumn
{1}{c}{1980}\\[6pt]
& & \multicolumn{1}{c}{Logistic--normal} & \multicolumn
{1}{c@{}}{Combined}\\[6pt]
Intercept treatment A & $\xi_{0}$ & -1.6299\ (0.4354) & -1.6042\
(4.0263) \\
Slope treatment A & $\xi_{1}$ & -0.4042\ (0.0460) & -6.4783\
(1.4386) \\
Intercept treatment B & $\xi_{2}$ & -1.7486\ (0.4478) & -16.2079\
(3.5830)\\
Slope treatment B & $\xi_{3}$ & -0.5634\ (0.0602) & -8.0745\
(1.5997) \\
Std. dev. random effect & $\sqrt{d}$ & 4.0150\ (0.3812)&60.8835\
(14.2237) \\
Ratio & $\alpha/\beta$ &\multicolumn{1}{c}{---} & 0.2805\ (0.0350) \\
$-$2log-likelihood &&\multicolumn{1}{c@{}}{1248}&\multicolumn
{1}{c}{1240}\\
\hline
\end{tabular*}
\end{table*}
%
%t5 ###
\begin{table*}[b]
\tabcolsep=0pt
\tablewidth=417pt
\caption{Asthma study. Parameter estimates and standard
errors for the regression coefficients in (1) the exponential model,
(2) the exponential--gamma model, (3) the exponential--normal model  and
(4) the combined model. Estimation was done by maximum likelihood using
numerical integration over the normal random effect, if present}\label
{asthma estimates}
\begin{tabular*}{417pt}{@{\extracolsep{\fill}}lcd{2.12}d{2.12}@{}}
\hline
& & \multicolumn{2}{c@{}}{\textbf{Estimate (s.e.)}}\\
\cline{3-4}
\textbf{Effect} & \textbf{Parameter} & \multicolumn{1}{c}{\textbf{Exponential}} & \multicolumn
{1}{c@{}}{\textbf{Exponential--gamma}}    \\
\hline
Intercept & $\xi_{0}$ & -3.3709\ (0.0772) & -3.9782\ (15.354)\\
Treatment effect & $\xi_{1}$ & -0.0726\ (0.0475) & -0.0755\ (0.0605) \\
Shape parameter & $\lambda $ & 0.8140\ (0.0149) & 1.0490\ (16.106) \\
Std. dev. random effect & $\sqrt{d} $ & \multicolumn
{1}{c}{---}&\multicolumn{1}{c}{---} \\
Gamma parameter & $\gamma $ & \multicolumn{1}{c}{---}& 3.3192\
(0.3885) \\
$-$2log-likelihood &&\multicolumn{1}{c}{18,693}&\multicolumn
{1}{c@{}}{18,715}\\[6pt]
& & \multicolumn{1}{c}{Exponential--normal} & \multicolumn
{1}{c@{}}{Combined}\\[6pt]
Intercept & $\xi_{0}$ & -3.8095\ (0.1028) & 3.9923\ (20.337)\\
Treatment effect & $\xi_{1}$ & -0.0825\ (0.0731) & -0.0887\ (0.0842)\\
Shape parameter & $\lambda $ & 0.8882\ (0.0180) & 0.8130\ (16.535)\\
Std. dev. random effect & $\sqrt{d} $ & 0.4097\ (0.0386) & 0.4720\
(0.0416)\\
Gamma parameter & $\gamma $ & \multicolumn{1}{c}{---}& 6.8414\
(1.7146)\\
$-$2log-likelihood &&\multicolumn{1}{c@{}}{18,611}&\multicolumn
{1}{c}{18,629}\\
\hline
\end{tabular*}
\end{table*}

%s6.3 ###
\subsection{Recurrent Asthma Attacks in Children}\label{analysissurvival}

We will analyze the times-to-event, introduced in Section~\ref
{datasurvival}. We consider an exponential model, that is, a model of
the form (\ref{weibull1}) with $\rho=1$, and further a predictor of
the form
\[
\kappa_{ij}=\xi_0+b_i+\xi_1 T_i,
\]
where $T_i$ is an indicator for treatment and $b_i\sim N(0,d)$. Results
from fitting all four models (with/\break without normal random effect;
with/without gamma random effect)
can be found in Table~\ref{asthma estimates}.
A formal assessment of the treatment effect from all four models is
given in Table~\ref{asthma tests}.
The treatment effect $\xi_1$ is stably identifiable in all four
models. As can be seen from Table~\ref{asthma tests}, the treatment
effects are similar in strengths, but including both random effects
reduces the evidence, relative to the exponential model. Needless to
say, too parsimonious an association structure might lead to liberal
test behavior.

%s6.4 ###
\subsection{The Need for the Combined Model}

We have some evidence from the above three examples that there is a
need for the combined model. Some indication came, for example, from
the correlation functions in the epilepsy case. It is useful to perform
formal comparison of all nested models, using Wald statistics, for each
of the three cases. A summary is given in Table~\ref{table tests}.
Note that, owing to the familiar boundary problem that occurs when
testing for variance components, mixtures of a $\chi^2_0$ and $\chi
^2_1$ were used, instead of the conventional  $\chi^2_1$ (Molenberghs
and Verbeke, \citeyear{Molenberghs2007}).
In all three case studies it is clear that: (1) independence is
strongly rejected in favor of both a model with normal random effects
or a model with conjugate random effects; (2) on top of one set of
random effects, there is a clear need for the other set as well, hence
providing very strong evidence for the proposed combined model. The
evidence is extremely convincing in all three cases.

%t6 ###
\begin{table}
\tablewidth=190pt
\caption{Asthma study. Wald test results for the
assessment of treatment effect}\label{asthma tests}
\begin{tabular*}{190pt}{@{\extracolsep{\fill}}lcc@{}}
\hline
\textbf{Model}&$\bolds{Z}$\textbf{-value}&$\bolds{p}$\textbf{-value}\\
\hline
Exponential &$-$1.5283&0.1264\\
Exponential--gamma &$-$1.1293&0.2588\\
Exponential--normal&$-$1.2480&0.2120\\
Combined &$-$1.0534&0.2921\\
\hline
\end{tabular*}
\end{table}

%t7 ###
\begin{table}[b]
\tabcolsep=0pt
\caption{All three case studies. Wald test results for
comparison of nested models}\label{table tests}
\begin{tabular*}{\columnwidth}{@{\extracolsep{\fill}}lcd{2.2}c@{}}
\hline
\textbf{Null model}&\textbf{Alternative model}&\multicolumn{1}{c}{$\bolds{Z}$\textbf{-value}}&$\bolds{p}$\textbf{-value}\\
\hline
Epilepsy study\\
\quad Poisson &Negative-binomial & 20.68&$<$0.0001\\
\quad Poisson &Poisson--normal & 6.27&$<$0.0001\\
\quad Negative-binomial&Combined & 6.10&$<$0.0001\\
\quad Poisson--normal&Combined & 11.66&$<$0.0001\\[6pt]
Onychomycosis study\\
\quad Logistic&Beta--binomial &17.91&$<$0.0001\\
\quad Logistic&Logistic--normal&10.53&$<$0.0001\\
\quad Beta--binomial&Combined &4.28 &$<$0.0001\\
\quad Logistic--normal&Combined&8.01 &$<$0.0001\\[6pt]
{Asthma study}\\
\quad Exponential&Exponential--gamma & 8.54&$<$0.0001\\
\quad Exponential&Exponential--normal &10.63&$<$0.0001\\
\quad Exponential--gamma&Combined &8.54 &$<$0.0001\\
\quad Exponential--normal&Combined &3.99 &$<$0.0001\\
\hline
\end{tabular*}
\end{table}

These findings, taken together, imply that the data exhibit, at the
same time, within-subject correlation and overdispersion.

%s7 ###
\section{Concluding Remarks}

In this paper we have argued that, rather than choosing between normal
and nonnormal random effects, the latter often of a gamma, beta or
other conjugate type, both can usefully be integrated together into a
single model, which we have termed the combined model. Our work builds
upon that of Molenberghs, Verbeke and Dem\'etrio (\citeyear{MolenberghsDem2007}), who brought
together normal random effects to induce association between repeated
Poisson data, and a gamma distributed random factor in the log-linear
predictor to fine tune the overdispersion. Their model produces the
standard negative-binomial and Poisson--normal models as special cases,
both when there are repeated measures as well as with univariate outcomes.

The current paper builds upon this work, not only by considering other
important cases, such as binary and time-to-event data and, for
completeness, also the normally distributed case, but, in particular,
by providing an encompassing framework around it. Wherever possible,
explicit expressions for the marginal joint distributions are derived,
as well as for marginal means, variances, covariances and moments in
general. This is possible in all cases, including the Poisson and
Weibull cases, but for binary data the logit links defies such a closed
form. However, we showed that a switch to the probit link does allow
for closed forms. The existence of these closed forms, producing
expressions for a variety of generalized linear mixed models as special
cases, has not been known to its fullest extent. We discuss their
implications for: (1) general understanding; (2) derived quantities
such as correlations, treatment effects, etc.; and (3) the construction
of parameter estimation and implementation.

For the binary case, we have exploited the logit-probit relationship to
derive probit-based closed-form approximations to the logit case. For
the Weibull situation, we have additionally generated a family of
distributions that encompass an entire collection of Cauchy-type distributions.

To make these developments possible in their fullest generality, we
have introduced \textit{strong conjugacy},\break    which comes down to a version
of the well-known conjugacy that is compatible with the additional
introduction of normal random effects.

In terms of estimation, we have focused on maximum likelihood
estimation. This can be done by integrating over the random effects,
either fully analytically, using the explicit expressions derived, or
by combining analytic and numeric techniques. The latter has been
implemented in the SAS procedure NLMIXED, for the Poisson, binary and
survival cases, and applied to three case studies.

Of course, with the considerations of not only one but multiple sets of
random effects comes the obligation to reflect on the precise nature of
such latent structures. As underscored by Verbeke and Molenberghs
(\citeyear{Verbeke2009}), full verification of the adequacy of a random-effects structure
is not possible based on statistical considerations alone, because
there is a many-to-one map from hierarchical models to the implied
marginal model. Of course, this should not stop the user from
considering such models, but rather issues a word of caution.

A number of topics have been mentioned in this paper that deserve
further research. These include, but are not limited to, the following:
(1) the construction of model building and goodness-of-fit tool; (2) a
detailed study of the relative merits of various estimation methods and
their implementation; (3) a study of the identifiability of
(random-effects) parameters in the combined model; (4) the
incorporation of censoring in the survival case; and (5) the explicit
consideration of data types and models not considered here.

The Poisson, binary and Weibull cases have\break been implemented in the SAS
procedure\break NLMIXED. All datasets, programs and outputs can be found in a
WinZip archive on the web site\break \url{www.censtat.be/software}.

\section*{Acknowledgments}
 Financial support from the IAP research network
\#P6/03 of the Belgian Government (Belgian Science Policy) is
gratefully acknowledged. This work was partially supported by a grant
from Coordenadoria para o Aperfei\c{c}oamento de Pessoal de Nvel
Superior (CAPES), Brazilian science funding agency.

\begin{supplement}
\stitle{A family of generalized linear models for repeated measures
with normal and conjugate random effects: Calculation details\\}
\slink[doi]{10.1214/10-STS328SUPP}
\sdatatype{.pdf}
\sdescription{In Section A, generic approximate calculations are
provided.
Closed-form calculations for various cases are offered as well: for
the Poisson case
(Section B), for the binary case with logit link (Section~C), for the
binary case with probit
link (Section~D), and for the time-to-event case (Section E). Finally,
Section F is dedicated
to the derivation of marginal correlation functions.}
\end{supplement}

\end{document}